\documentclass[a4paper,10pt]{article}
\usepackage[utf8]{inputenc}
\usepackage{times}
\usepackage{setspace}
\usepackage{url}
\usepackage{geometry}
\usepackage{graphicx}
\usepackage{subfigure}
\usepackage{float}
\usepackage{caption}
\usepackage{epstopdf}
\usepackage{epsfig}
\usepackage{dcolumn}
\usepackage{amsthm,amsmath}
\usepackage{amssymb}
\usepackage{fancyhdr}
\usepackage{tikz}
\usetikzlibrary{calc,fadings,decorations.pathreplacing}
\usepackage{verbatim}

\newcommand\pgfmathsinandcos[3]{%
  \pgfmathsetmacro#1{sin(#3)}%
  \pgfmathsetmacro#2{cos(#3)}%
}
\newcommand\LongitudePlane[3][current plane]{%
  \pgfmathsinandcos\sinEl\cosEl{#2} 
  \pgfmathsinandcos\sint\cost{#3} 
  \tikzset{#1/.estyle={cm={\cost,\sint*\sinEl,0,\cosEl,(0,0)}}}
}
\newcommand\LatitudePlane[3][current plane]{%
  \pgfmathsinandcos\sinEl\cosEl{#2} 
  \pgfmathsinandcos\sint\cost{#3} 
  \pgfmathsetmacro\yshift{\cosEl*\sint}
  \tikzset{#1/.estyle={cm={\cost,0,0,\cost*\sinEl,(0,\yshift)}}} %
}
\newcommand\DrawLongitudeCircle[2][1]{
  \LongitudePlane{\angEl}{#2}
  \tikzset{current plane/.prefix style={scale=#1}}
  \pgfmathsetmacro\angVis{atan(sin(#2)*cos(\angEl)/sin(\angEl))} %
  \draw[current plane] (\angVis:1) arc (\angVis:\angVis+180:1);
  \draw[current plane,dashed] (\angVis-180:1) arc (\angVis-180:\angVis:1);
}
\newcommand\DrawLatitudeCircle[2][1]{
  \LatitudePlane{\angEl}{#2}
  \tikzset{current plane/.prefix style={scale=#1}}
  \pgfmathsetmacro\sinVis{sin(#2)/cos(#2)*sin(\angEl)/cos(\angEl)}
  \pgfmathsetmacro\angVis{asin(min(1,max(\sinVis,-1)))}
  \draw[help lines,current plane] (\angVis:1) arc (\angVis:-\angVis-180:1);
  \draw[help lines,current plane,dashed] (180-\angVis:1) arc (180-\angVis:\angVis:1);
}
\newcommand\DrawLatitudeCircleL[2][1]{
  \LatitudePlane{\angEl}{#2}
  \tikzset{current plane/.prefix style={scale=#1}}
  \pgfmathsetmacro\sinVis{sin(#2)/cos(#2)*sin(\angEl)/cos(\angEl)}
  \pgfmathsetmacro\angVis{asin(min(1,max(\sinVis,-1)))}
  \draw[current plane] (\angVis:1) arc (\angVis:-\angVis-180:1);
  \draw[current plane,dashed] (180-\angVis:1) arc (180-\angVis:\angVis:1);
}

\tikzset{%
  >=latex, 
  inner sep=0pt,%
  outer sep=2pt,%
  mark coordinate/.style={inner sep=0pt,outer sep=0pt,minimum size=3pt,
    fill=black,circle}%
}

\begin{large}
\author{Shivraj Prajapat \footnote{Indian Institute of Science Education and Research(IISER), Pune 411008, India ,Email: shraprajapat@gmail.com}~~$^{,\dagger}$ ~,~Yendrembam Chaoba Devi \footnote{S.N.Bose National Centre For Basic Sciences, Salt Lake, Kolkata 700098, India, Email: biswajit@bose.res.in, chaoba@bose.res.in}~, Aritra K. Mukhopadhyay \footnote{Indian Institute of Science Education and Research(IISER) Kolkata, Nadia 741 252 WB, India, Email: aritra1910@gmail.com}\\ Biswajit Chakraborty$^{\dagger}$ ,~~ Frederik G Scholtz \footnote{National Institute for Theoretical Physics (NITheP), Stellenbosch 7602, South Africa, Email: fgs@sun.ac.za}
}
\end{large}

\begin{LARGE}
\title{Connes Distance function on fuzzy sphere and the connection between geometry and statistics}
\end{LARGE}

\begin{document}
\maketitle

\begin{abstract}
An algorithm to compute Connes spectral distance, adaptable to the Hilbert-Schmidt operatorial formulation of non-commutative quantum mechanics, was developed earlier by introducing the appropriate spectral triple and used to compute infinitesimal distances in the Moyal plane, revealing a deep connection between geometry and statistics. In this paper, using the same algorithm, the Connes spectral distance has been calculated in the Hilbert-Schmidt operatorial formulation for the fuzzy sphere whose spatial coordinates satisfy the $su(2)$ algebra. This has been computed for both the discrete, as well as for the Perelemov's $SU(2)$ coherent state. Here also, we get a connection between geometry and statistics which is shown by computing the infinitesimal distance between mixed states on the quantum Hilbert space of a particular fuzzy sphere, indexed by $n\in\mathbb{Z}/2$.
\end{abstract}

\section{Introduction}
  At the Planck length scale, it is generally believed that the continuity of space-time breaks down and we have a fuzzy space-time. This implication comes from consideration of both quantum and gravity effects at this scale \cite{b9}, which was corroborated later through a low energy effect of string theory \cite{b10}. Depending upon the structure of non-commutativity between space-time coordinates or for that matter between the spatial coordinates, (if the time is taken to be a c-number parameter, for simplicity), one can define the non-commutative space in many ways \cite{b8}. Some fuzzy spaces which are simpler and have been widely studied are as follows:
  \begin{enumerate}
   \item{Moyal Space:-  \begin{equation}
                          [\hat{x}_i , \hat{x}_j ] = i\theta_{ij},~~i,j=1,2,3~~\text{and}~~\theta_{ij} ~\text{is an anti-symmetric constant matrix} 
                          \end{equation}}
  \item{Fuzzy Sphere:- \begin{equation}
                          [\hat{x}_i , \hat{x}_i ] = i\lambda\epsilon_{ijk}\hat{x}_k, ~~~\lambda~\text{is some parameter and}~ \epsilon_{ijk} ~ \text{is the usual anti-symmetric tensor} \label{fuzzycom1}
                          \end{equation}}                        
  \end{enumerate}
The notion of distance on such fuzzy spaces can be given by introducing a spectral triple ($\mathcal{A},\mathcal{H},\mathcal{D}$)  introduced by Connes \cite{b4} to define geometry for a general space (including discrete space). 

The fuzzy space of the first type has been extensively studied using both the star product and the Hilbert-Schmidt operator method that was initiated in \cite{b14}. The infinitesimal distances on such spaces have been calculated in \cite{b11}, making use of the Moyal star product, and in \cite{b1} using the Hilbert-Schmidt operator method. The advantage of the Hilbert-Schmidt operatorial formulation is that it can bypass the use of any star product and the resulting ambiguities \cite{b15,b16}.  As has been shown in \cite{b17} the different star products (Moyal and Voros) stem from different choices of bases, which lead to different classes of function when states are represented in these bases.  This implies that equivalence between different choices of star products cannot be guaranteed, especially in path integral formulations,  without taking due care of this point.  In addition the Moyal product does not admit a position POVM, unlike the Voros product.  The latter therefore admits a completely consistent quantum interpretation of position measurement as a weak measurement, while the former does not.  These issues, expansions of them and possible interpretations thereof can be found in   \cite{b18}. 

 In \cite{b1}, an algorithm adaptable for this Hilbert-Schmidt operatorial formulation was developed and has been applied to calculate the infinitesimal distance on the 2D Moyal plane, both in the classical and quantum Hilbert space, revealing a deep connection between geometry and statistics, which is in turn intimately connected to fact that meaning can only be given to a position measurement in the weak sense, implying an inherent ignorance of the state of a particle even after a position measurement.  This can again be interpreted in the context of additional degrees of freedom that cannot be probed in a position measurement \cite{b18} . The purpose of this paper is to show that this algorithm can also be applied on the second type of fuzzy space, whose Hilbert-Schmidt operator formulation was developed in \cite{b12}, to analyze the Coulomb problem in this kind of non-commutative space.
  
The second type of fuzzy space is realized in the presence of magnetic field produced by magnetic monopole \cite{b3}. Such type of non-commutativity is found between operators which are like the angular momentum operators $\hat{J}_i$ in quantum mechanics and for which  we can define the simultaneous eigenstates $\rvert j,m\rangle$ of the Casimir operator $\hat{\vec{J}}^{2}$ and $\hat{J}_{3}$. In the same way, we can define the simultaneous eigenstates $\rvert n,n_{3}\rangle$ of radius squared operator $\hat{\vec{x}}^2$ and $\hat{x}_3$ for the above type of fuzzy space (\ref{fuzzycom1}). The notion of 3D configuration space is now replaced with the Hilbert space spanned by the kets $\rvert n,n_{3}\rangle$. The eigenvalues of the radius-squared operator are of the form $\lambda^{2}n(n+1) , n\in \mathbb{Z}/2$ and the eigenvalues of $\hat{x}_{3}$ are $\lambda n_{3}$ where $ -n\leq n_{3} \leq n$. The radius is thus quantized and each sphere with the fixed radius is referred to as fuzzy sphere. In this paper, we will 
try to give the 
notion of Connes infinitesimal distance on the fuzzy sphere using the same prescription given in \cite{b1}. It therefore provides an alternative perspective to the one in \cite{b2}.
  
The paper is organized in the following way. In section 2, we revisit the Hilbert-Schmidt operator formulation of the fuzzy sphere \cite{b12} and we revisit the construction of Dirac operator on $S^3$ and hence on $S^2$ \cite{b6} in section 3. This is the Dirac operator we need in order to define a spectral triple on the fuzzy sphere. With this, we give the appropriate spectral triple to calculate the Connes distance function on the fuzzy sphere in section 4. In this section, we obtain the infinitesimal Connes distance function on the configuration space. Also in a subsection, we revisit the construction of Perelemov coherent states on $S^2$ \cite{b5} and calculate the infinitesimal distance between these coherent states. In section 5, we give the spectral triple for the quantum Hilbert space and calculate the distance function for both pure and mixed states. The distance between mixed states shows the connection between geometry and statistics. Finally, we provide some relevant calculations in the appendix.

\section{Configuration space and Quantum Hilbert space of the fuzzy sphere}

For the case of fuzzy spheres, we have the spatial non-commutativity of Lie algebra type where the space coordinates satisfy the su(2) algebra:
\begin{equation}
[\hat{x}_{i},\hat{x}_{j}] = i\lambda\epsilon_{ijk}\hat{x}_{k} , 
\end{equation}
with $\lambda$ a non-commutative parameter of length dimension.

Now, let us define the creation and annihilation operators of two independent harmonic oscillators, $\hat{\chi}^\dagger_\alpha$ and $\hat{\chi}_\alpha$ with the following commutators between them
 \begin{equation}
[\hat{\chi}_{\alpha},\hat{\chi}^{\dagger}_{\beta}] = \frac{\lambda}{2}\delta_{\alpha\beta}~,~~[\hat{\chi}_{\alpha},\hat{\chi}_{\beta}] = 0 = [\hat{\chi}^{\dagger}_{\alpha},\hat{\chi}^{\dagger}_{\beta}]~;~~~~~~\alpha,\beta=1,2 \label{c10}
\end{equation}
such that we have the Jordan-Schwinger map between the $\hat{x}_i$ and $\hat{\chi}$'s:  
\begin{equation}
 \hat{x}_{i} =  \hat{\chi}^{\dagger}\sigma_{i}\hat{\chi} =  \hat{\chi}^{\dagger}_{\alpha}\sigma_{i}^{\alpha\beta}\hat{\chi}_{\beta},\label{JSmap}
\end{equation}
where $\sigma_i$ are the Pauli matrices.

Labeling the two harmonic oscillators by $n_1$ and $n_2$, we have the eigenvalue equation of the number operator $\hat{N}=\hat{\chi}_\alpha^\dagger\hat{\chi_\alpha}$ as,
\begin{equation}
 \hat{N}\rvert n_{1},n_{2}\rangle=\hat{\chi}_\alpha^{\dagger}\hat{\chi}_\alpha \rvert n_{1},n_{2}\rangle = \frac{\lambda}{2}(n_{1}+n_{2})\rvert n_{1},n_{2}\rangle,
\end{equation}
 where  \begin{equation}
\rvert n_{1}, n_{2}\rangle =  \sqrt{\frac{(2/\lambda)^{n_{1}+n_{2}}}{n_{1}! n_{2}!}}\chi^{\dagger n_{1}}_{1}\chi^{\dagger n_{2}}_{2}\rvert 0\rangle.
\end{equation}
This is also the simultaneous eigenstate of the radius squared operator $\hat{\vec{x}}^{2}$ and $\hat{x}_3$, i.e.
\begin{eqnarray}
\hat{\vec{x}}^{2}\rvert n_{1},n_{2}\rangle  &=& \lambda^{2}n(n+1)\rvert n_{1},n_{2}\rangle ~~,~~n= \frac{n_{1}+n_{2}}{2} \label{r^2no.}, \\
\hat{x}_{3}\rvert n_{1},n_{2}\rangle &=& \lambda n_{3}\rvert n_{1},n_{2}\rangle ~~~~,~~~~~ n_{3}=\frac{n_{1}-n_{2}}{2}.
\end{eqnarray}
The state $\rvert n_{1},n_{2}\rangle$ can alternatively be relabeled by $\rvert n , n_{3}\rangle$, where $n, n_{3}~\in \mathbb{Z}/2$ and $-n\leq n_3 \leq n$. The ladder operators ~ $\hat{x}_{\pm} = \hat{x}_{1}\pm i\hat{x}_{2}$ ~ satisfy 
\begin{eqnarray}
[\hat{x}_{3},\hat{x}_{\pm}]= \pm\lambda \hat{x}_{\pm} ~~~~~~,~~~~~
[\hat{x}_{+},\hat{x}_{-}]&=& 2\lambda \hat{x}_{3}
\end{eqnarray}
so that,
\begin{eqnarray}
\hat{x}_{\pm}\rvert n,n_{3}\rangle =\lambda \sqrt{n(n+1)-n_{3}(n_{3}\pm1)}\rvert n,n_{3}\pm1\rangle.\\
\end{eqnarray}
The 3D configuration space i.e. the classical Hilbert space $\mathcal{F}_c$  is given by  
\begin{eqnarray}
\mathcal{F}_c = Span\{\rvert n,n_{3}\rangle\} \label{confspace},
\end{eqnarray}
where the radius is quantized as in (\ref{r^2no.}). Each $n$ corresponds to the fixed sphere of radius $\lambda\sqrt{n(n+1)}.$ For fixed $n$ the  Hilbert space is restricted to (2n+1)-dimensional sub-space
\begin{eqnarray}
\mathcal{F}_{n} = Span\{\rvert n,n_{3}\rangle ~~\rvert~~ \text{n is fixed}, -n \leq n_{3}\leq n\}. \label{c20}
\end{eqnarray}
The Hilbert-Schmidt operators  which act on the $\mathcal{F}_{c}$  can be written as 
\begin{equation}
\Psi \in Span\{\rvert n,n_{3}\rangle\langle n',n'_{3}\rvert\}.
\end{equation}
The quantum Hilbert space, in which the physical states are represented, consists, however, of those operators generated by coordinate operators only and, since these commute with the Casimir, the elements of the quantum Hilbert space must in addition commute with the Casimir, i.e., must be diagonal in $n$.  The quantum Hilbert space $\mathcal{H}_q$ of the fuzzy space of type (\ref{fuzzycom1}) is therefore 
\begin{equation}
 \mathcal{H}_q=\{\Psi \in Span\{\rvert n,n_{3}\rangle\langle n,n'_{3}\rvert\}:~~\text{tr}_c(\Psi^\dag\Psi)<\infty\}.
\end{equation}
Note that the quantum Hilbert space is the direct sum of the subspaces $\mathcal{H}_n$ with a fixed value of $n$:
\begin{equation}
 \mathcal{H}_n=\{\Psi \in Span\{\rvert n,n_{3}\rangle\langle n,n'_{3}\rvert\}\equiv |n_3,n'_3):~~\text{tr}_c(\Psi^\dag\Psi)<\infty~~ \text{with fixed}~~ n\}. \label{hn}
\end{equation}
Note that we have chosen a compact notation $ |n_3,n'_3) \in \mathcal{H}_n $ by suppressing $n$. 

One can think of the fuzzy space of the second type (\ref{fuzzycom1}) as the space of many fuzzy spheres with different radii labeled by $n$ and on each fuzzy sphere we can vary $n_3$ from $-n$ to $+n$ such that one has the notion of two types of distances: one between the fuzzy spheres indexed by different $n$ and another between the generalized points on a particular fuzzy sphere characterized by different $n_3$ for a given $n$. Here, our intention is to define the Connes spectral distance between the generalized points on the fuzzy sphere. In order to define distance on the fuzzy space, we will use the formula for the generalized distance which is called the spectral distance. To define such generalized distance on the fuzzy space where the notion of the usual point is lost, we need to introduce an appropriate spectral triple  ($\mathcal{A},\mathcal{H},\mathcal{D}$) on the space of interest.  Here $\mathcal{A}$ is an involutive algebra (which will capture the topological information of the space upon which it is defined) 
acting on the Hilbert space $\mathcal{H}$ through an appropriate representation and $\mathcal{D}$ is the Dirac operator which is a self adjoint operator acting on the same $\mathcal{H}$.

The Connes spectral distance between the generalized points, which are given by the pure states of the algebra $\mathcal{A}$, is given by
 \begin{equation}
d_\mathcal{D}(\omega, \tilde{\omega}) \stackrel{.}{=} \sup_{a\in \mathcal{A}}\big\{|\omega(a)-\tilde{\omega}(a)|, \lVert[\mathcal{D},\pi(a)]\rVert\leq 1\big\} \label{sp.dist.}
\end{equation}
where $\pi$ denotes the representation of $\mathcal{A}$ on $\mathcal{H}$ and $\omega$ denotes a pure state of $\mathcal{A}$. 

In order to construct the spectral triple on the fuzzy sphere we need to find the Dirac operator acting on a relevant Hilbert space of the fuzzy sphere. For this purpose, it will be better to go through the construction of Dirac operator on $S^3$ and $S^2$ first and then proceed for the fuzzy sphere.

\section{Construction of Dirac operators on $S^{3}$ and $S^{2}$}

In this section we briefly review the construction of Dirac operator for commutative 2 and 3-spheres i.e. $S^{2}$ and $S^{3}$.  This construction then will pave the way to generalize it for non-commutative case. For this , we essentially follow \cite{b6}.

We begin by introducing the flat  $C^{2}_0$ manifold which is isomorphic to $R^{4}-\{0\}$ and is defined through the set of nonzero complex doublets:
\begin{equation}
C^{2}_0  = \{\chi = \left(
\begin{array}{c}
\chi_{1}\\
\chi_{2}\\
\end{array}
\right)\in C^{2}\rvert \chi \neq 0\}
\end{equation}
The  $CP^{1}$ manifold  consists of nonzero complex doublets with the identification 
\begin{equation}
\left(
\begin{array}{c}
\chi_{1}\\
\chi_{2}\\
\end{array}
\right) \sim \left(
\begin{array}{c}
\lambda\chi_{1}\\
\lambda\chi_{2}\\
\end{array}
\right) ;~~~ C \ni\lambda \neq 0, \label{c5}
\end{equation}
which are rays, i.e., the set of complex lines passing through the origin in $C^{2}$. The representative points on each line is chosen  by first imposing the restriction $\chi^{\dag}\chi = r= \textit{constant}$. 
\begin{equation}
 \{\chi = \left(
\begin{array}{c}
\chi_{1}\\
\chi_{2}\\
\end{array}
\right)\in C^{2}\ \rvert \chi^{\dagger}\chi= \rvert\chi_{1}\rvert^{2}+\rvert\chi_{2}\rvert^{2} = r \}, \label{c1}
\end{equation}
which therefore clearly is  $S^{3}$. With $r=1$ we have $S^{3}\sim SU(2).$  We then have still U(1) freedom left in the constraint in (\ref{c1}) with respect to the transformation
\begin{equation}
\chi_{i} \rightarrow e^{i\theta}\chi_{i}.
\end{equation}
If we choose  the gauge  from this  U(1) freedom,  we can identify $CP^{1}$ = $S^{2}$ =  SU(2)/U(1) i.e. choosing the section from the U(1) bundle over $S^{2}$. Two particular ways of doing this are :
\begin{enumerate}
\item{Choose a section from the U(1) bundle where $\chi_{1}$ is real- $\chi^{*}_{1} = \chi_{1}$ in the neighborhood $U_{+}:  \chi_{1}\neq 0$. With this the constraint in (\ref{c1}) reduces to 
\begin{equation}
\chi_{1}^{2} ~+~ \rvert\chi_{2}\rvert^{2} = r \label{c3}
\end{equation}
so that this can be clearly identified with  $S^{2}$. This can  equivalently be obtained by the transformation 
\begin{equation}
U_{+}: \chi\rightarrow \chi' =\left(
\begin{array}{c}
\chi'_{1}\\
\chi'_{2}\\
\end{array}
\right) = \left(\frac{\chi^{*}_{1}}{\chi_{1}}\right)^{1/2}\left(
\begin{array}{c}
\chi_{1}\\
\chi_{2}\\
\end{array}
\right) = \left(\frac{\chi^{*}_{1}}{\chi_{1}}\right)^{1/2}\chi.
\end{equation}
 Now choosing $\lambda = 1/\chi_{1}$ in $(\ref{c5})$ we can write
\begin{equation}
\chi =  \left(
\begin{array}{c}
\chi_{1}\\
\chi_{2}\\
\end{array}
\right) \approx \left(
\begin{array}{c}
1\\
\rho\\
\end{array}
\right) ~~;~~ \rho = \chi_{2}/\chi_{1}, \label{zzz}
\end{equation}
where $\rho$ is a complex number representing the  stereographic projection of a point of the sphere from the south pole. Thus $U_{+}$ corresponding to the northern hemisphere which is parametrized by $\rho$}.
\item{Like-wise, choose a section from the U(1) bundle where $\chi_{2}$ is real- $\chi_{2} = \chi^{*}_{2}$ in the neighborhood $U_{-}: \chi_{2} \neq 0$. With this the constraint in (\ref{c1}) reduces to -
\begin{equation}
\rvert\chi_{1}\rvert^{2}+\chi_{2}^{2} = r, \label{c4}
\end{equation}
which again clearly is $S^{2}.$
 Here also similar to the previous case the doublet can be written as 
\begin{equation}
 \chi= \left(
\begin{array}{c}
\chi_{1}\\
\chi_{2}\\
\end{array}
\right)\approx\left(
\begin{array}{c}
\eta\\
1\\
\end{array}
\right)~~ ;~~ \eta = \chi_{1}/\chi_{2}, \label{yyy}
\end{equation}
where $\eta$ is a complex number representing the  stereographic projection of a point of the sphere from the north pole. Thus $U_{-}$ corresponds to the southern hemisphere.}
\end{enumerate}

Both the chart $U_{+}$ and $U_{-}$ is interpreted as covering the northern hemisphere and southern hemisphere respectively. The chart $U_{+}$ misses only the south pole and the chart $U_{-}$ misses only the north pole. In the overlapping region $U_{+}\bigcap U_{-}$, one therefore has $\eta = 1/\rho$. For the unit sphere $S^{2}$, the transition function relating sections $\psi_{\pm}$ of the tautological line bundle thus takes the form:
\begin{eqnarray}
 \psi_{+} = e^{i\varphi}\psi_{-}. \label{A1}
\end{eqnarray}
The Cartesian coordinates of the $S^{2}$ in the respective chart is defined through the Hopf map 
\begin{eqnarray}
x'_{i} = \chi'^{\dag}\sigma_{i}\chi' \in U_{+} , i= 1,2,3, \label{shiv10} \\
x''_{i} = \chi'^{\dag}\sigma_{i}\chi' \in U_{-} , i= 1,2,3, \label{shiv1}
\end{eqnarray}
where $\sigma_i$ are the Pauli matrices. In the region $U_{+}\cap U_{-}$ the coordinates $x'_{i}$ and $x''_{i}$ coincide i.e. $x'_{i} = x''_{i} = x$, as they are gauge invariant. Note that the Jordan-Schwinger map (\ref{JSmap}) is just the operatorial version of this Hopf map.

Now the $\chi\in S^{3}$ can be parametrized in terms of Euler's angle  as
\begin{eqnarray}
\chi_1 = r^{\frac{1}{2}}\cos\frac{\theta}{2}e^{\frac{i}{2}(\varphi+\psi)}, \label{rr2}\\
\chi_2 = r^{\frac{1}{2}}\sin\frac{\theta}{2}e^{-\frac{i}{2}(\varphi-\psi)}, \label{f1}
\end{eqnarray} 
so that one gets the familiar coordinates of $S^2(\vec{x}^2=r^2)$ as $x_{1} = r\sin\theta\cos\varphi$, $x_{2}=r\sin\theta\sin\varphi$ and $x_{3} =r\cos\theta)$. Further this sphere $\vec{x}^2=r^2$ is distinct from the ones occurring in (\ref{c3}, \ref{c4}).

The $\chi' , \chi''\in S^{2}$- the sections corresponding to $U_{\pm}$ are then obtained as 
\begin{eqnarray}
\chi' = e^{-\frac{i}{2}(\psi+\varphi)}\chi,  \label{QQ1}\\
\chi'' = e^{\frac{i}{2}(\varphi-\psi)}\chi.  \label{QQ2}
\end{eqnarray}
In the region $U_{+}\cap U_{-}$ we have the transition map to go from one chart to other.
\begin{equation}
\chi' = e^{i\varphi}\chi'' ,  \label{f9}
\end{equation}
 which clearly has the same form as (\ref{A1}). Before, we construct the spinor-bundle, let us revisit (\ref{zzz}) where $\rho$ represents a point in the complex plane that coordinates a neighbourhood of  $CP^{1}$. Now if we consider a map $\rho\rightarrow\rho^{n}$ (say), with $n$ being a positive integer, then it is clear that $\rho^{n}$ winds around the complex plane $n$ times in the counter-clock-wise direction. Like-wise, $\eta$-occurring in (\ref{yyy}) will also wind around $n$ times, but in the clock-wise direction. They are just other way around for the corresponding  complex conjugate variables $\rho^{*}$ and $\eta^{*}$ , as one can easily see. Consequently considering any function
 \begin{eqnarray}
 f(\rho,\rho^{*}) = (\rho^{*})^{m_{1}}\rho^{m_{2}}  \label{A2}
 \end{eqnarray}
 given in terms of the monomial, and defined in the region $U_{+}$, the total winding number is $m_{2}-m_{1}$, which is nothing but the algebraic sum of the respective winding numbers. For generic cases this function can not be regarded as a section in the tautological line-bundle, rather on other non-canonical U(1) bundle over $S^{2}$, with the associated transition function $e^{ik\varphi}$ \cite{b7}
 \begin{eqnarray}
  \psi_{+} = e^{ik\varphi}\psi_{-}. \label{A3}
 \end{eqnarray}
It is only for $k=1$, that one can identify the bundle to be  $S^{3}$. With this background let us now consider first the
 spinor bundle over the $C_{0}^{2}$ manifold whose sections are of the form
\begin{equation}
 \Psi = \left(
\begin{array}{c}
\psi_{1}(\chi_{\alpha},\chi^{*}_{\beta})\\
\psi_{2}(\chi_{\alpha},\chi^{*}_{\beta})\\
\end{array}
\right)~~~~ \alpha,\beta = 1,2 ,\label{c7}
\end{equation}
 where $\psi_{\alpha}$ are polynomials i.e. homogeneous functions of  $\chi_{\alpha} $ and $\chi_{\alpha}^{*}$:
\begin{equation}
\psi_{\alpha} = \sum_{m_{1},m_{2},n_{1},n_{2}}C^{\alpha}_{m_{1},m_{2},n_{1},n_{2}}(\chi^{*}_{1})^{m_{1}}(\chi^{*}_{2})^{m_{2}}\chi^{n_{1}}_{1}\chi^{n_{2}}_{2} ~~,~~ m_{1},m_{2},n_{1},n_{2} \in Z.    \label{f2}
\end{equation}
Note that we have used here the homogeneous coordinates $\chi_{\alpha},\chi^{*}_{\alpha}$ rather than the stereographic variables $\rho$ or $\eta$ in the argument of both the functions $\psi_{1}$ and $\psi_{2}$. These can be regarded  as a doublet of scalar fields like (\ref{A2}), but transforming under SU(2). This is a trivial bundle, defined globally on $C^{2}_{0}$. The section of the spinor bundle over $S^{3}$ is obtained by imposing  restrictions on $\chi s$ of the form 
\begin{equation}
\chi^{\dagger}\chi = r, \label{c6}
\end{equation}
and the sections of the spinor bundle over $S^{2}$ is defined using $\chi'$ in $U_{+}$ and $\chi''$ in $U_{-}$(\ref{QQ1},\ref{QQ2}).

The differential operator 
\begin{equation}
J_{i} = \frac{1}{2}(\chi_{\alpha}\sigma^{\beta\alpha}_{i}\partial_{\chi_{\beta}}-\chi^{*}_{\alpha}(\sigma^{*})^{\beta\alpha}_{i}\partial_{\chi^{*}_{\beta}}) \label{c15}
\end{equation}
acts on the spinors (\ref{c7}) defined on $C_{0}^{2}$. Since these differential operators are independent of $r$ (see (\ref{f4}) below)and  depends upon $\theta$~,~$\varphi$,~and $\psi$, these operators therefore act on $S^{3}$ as well  and can be identified with the orbital part of the rotation generators for the spinors on $S^{3}$ (See appendix A1). Similarly the dilatation  operator, is given by the differential operator
\begin{equation}
K = \frac{1}{2}(\chi_{\alpha}\partial_{\chi_{\beta}}-\chi^{*}_{\alpha}\partial_{\chi^{*}_{\beta}}). \label{c16}
\end{equation}
Since each term in (\ref{f2}) can be factored in to homogeneous holomorphic and anti-holomorphic functions, $K$ yields, by Euler's theorem, the net winding number.  The operator $J_{i}$ and $K$ can be expressed in terms of Euler's angle, using  (\ref{rr2},\ref{f1}),
\begin{eqnarray}
J_{1} &=& i\sin\varphi\frac{\partial}{\partial\theta}+i\cos\varphi\cot\theta\frac{\partial}{\partial\varphi}-i\frac{\cos\varphi}{\sin\theta}\frac{\partial}{\partial\psi} ,\nonumber \\
J_{2} &=& -i\cos\varphi\frac{\partial}{\partial\theta}+i\cot\theta\sin\varphi\frac{\partial}{\partial\varphi}-i\frac{\sin\varphi}{\sin\theta}\frac{\partial}{\partial\psi}, \nonumber\\        
J_{3} &=& -i\frac{\partial}{\partial\varphi}, \nonumber \\ \label{shiv}
K &=& i\frac{\partial}{\partial\psi}. \label{f4}
\end{eqnarray}
The operators $J_{i}$  satisfy SU(2) algebra 
\begin{equation}
 [J_{i},J_{j}] = i\epsilon_{ijk}J_{k}.
\end{equation}
Since the component of the sections of the spinor bundle on the unit ($r=1$), $S^{3}$, is of the form (\ref{f2}), it can be written in terms of the Euler's angles introduced in (\ref{rr2},\ref{f1}), 
\begin{eqnarray}
\psi_{\alpha} 
 = \sum_{m_{1},m_{2},n_{1},n_{2}}C^{\alpha}_{m_{1},m_{2},n_{1},n_{2}}(\cos\frac{\theta}{2})^{m_{1}+n_{1}}(\sin\frac{\theta}{2})^{n_{2}+m_{2}}\\
\times e^{-i\frac{\varphi}{2}(m_{1}-m_{2}-n_{1}+n_{2})}e^{-i\frac{\psi}{2}(m_{1}+m_{2}-n_{1}-n_{2})} . \label{f10}
\end{eqnarray}
The last exponential factor involving $\psi$ clearly indicates that the sections on $S(C^{2})$ and $S(S^{3})$ can be further divided into classes of different sub-bundles , indexed by the eigenvalue $k$ of the dilatation operator $K$ (\ref{f4}): $S_{k}(C^{2})$ and $S_{k}(S^{3})$ where 
\begin{equation}
k = m_{1}+m_{2}-n_{1}-n_{2}, \label{c8}
\end{equation}
is the eigenvalue of the operator $K$ (\ref{f4})
\begin{equation}
K\Psi = k\Psi.      \label{f8}
\end{equation}
The section of the spinor bundle over $S^{2}$ mentioned earlier ,\footnote{The section of the spinor bundle over $S^{2}$ can be obtained by making use of (\ref{rr2},\ref{f1}) by the gauge fixing i.e. by setting $ \psi=-\varphi $ in $ U_{+}$ and $\psi= \varphi $ in $U_{-}$. In both the charts the component of the section differs in the exponential factor in the form $e^{i\varphi(m_{2}-n_{2})}$ in $U_{-}$ and $e^{i\varphi(n_{1}-m_{1})}$ in $U_{+}$. If we have the condition k = 0, the components match in both the chart and $\Psi$ is globally defined on $S^{2}$ and we obtain a section on the trivial bundle. On the other hand, for  $k \neq 0$, then we have the transition rule to go from one chart to another in the form $e^{i\varphi(m_{2}-n_{2})}$ =  $e^{i\varphi(n_{1}-m_{1}-k)}$, the k is called the winding number, introduced earlier and is the eigenvalue of the K operator (\ref{f8}) }
 has the form
\begin{equation}
\Psi' = \left(
\begin{array}{c}
\psi'_{1}(\chi'_{\alpha},\chi'^{*}_{\beta})\\
\psi'_{2}(\chi'_{\alpha},\chi'^{*}_{\beta})\\
\end{array}
\right)~~~~ \alpha,\beta = 1,2 ~~~ on~~~ U_{+},
\end{equation}
\begin{equation}
\Psi'' = \left(
\begin{array}{c}
\psi''_{1}(\chi''_{\alpha},\chi''^{*}_{\beta})\\
\psi''_{2}(\chi''_{\alpha},\chi''^{*}_{\beta})\\
\end{array}
\right)~~~~ \alpha,\beta = 1,2 ~~~~on ~~~~U_{-}.
\end{equation}
In terms of Euler's angle they can be  written as
\begin{eqnarray}
\Psi'_{\alpha} &=& \sum a^{\alpha}_{n_{1}, n_{2},m_{1},m_{2}}\chi'^{*m_{1}}_{1}\chi'^{*m_{2}}_{2}\chi'^{n_{1}}_{1}\chi'^{n_{2}}_{2}  ~~~ \text{in} ~~~U_{+}  \\
& =& \sum a^{\alpha}_{n_{1}, n_{2},m_{1},m_{2}}r^{\frac{m_{1}+m_{2}+n_{1}+n_{2}}{2}}(\cos\frac{\theta}{2})^{m_{1}+n_{1}}(\sin\frac{\theta}{2})^{m_{2}+n_{2}}e^{i\varphi(m_{2}-n_{2})},\nonumber \\
\Psi''_{\alpha} &=& \sum a^{\alpha}_{n_{1}, n_{2},m_{1},m_{2}}\chi''^{*m_{1}}_{1}\chi''^{*m_{2}}_{2}\chi''^{n_{1}}_{1}\chi''^{n_{2}}_{2}  ~~~ \text{in} ~~~U_{-}  \\
& =& \sum a^{\alpha}_{n_{1}, n_{2},m_{1},m_{2}}r^{\frac{m_{1}+m_{2}+n_{1}+n_{2}}{2}}(\cos\frac{\theta}{2})^{m_{1}+n_{1}}(\sin\frac{\theta}{2})^{m_{2}+n_{2}}e^{i\varphi(n_{1}-m_{1})}. \nonumber
\end{eqnarray}
Since these are independent of $\psi$, they live on $S^{2}$. Further, as  the coefficients in both the expression are  same in the vicinity of equator ($\theta = \pi/2$) so that in $U_{+}\bigcap U_{-}$ we have
\begin{equation}
\Psi' = e^{ik\varphi+i\delta} \Psi'', 
\end{equation}
where $k$ (\ref{c8}) can now be identified with   the topological index (Chern class), and $\delta$ may be  constant or a globally defined function on $S^{2}.$ Consequently,  the two sections $\Psi_{0}$ and $\Psi_{1}$ on $S^{2}$ is called equivalent if 
\begin{equation}
\Psi'_{0}=\Psi'_{1}~~~ ,~~~  \Psi''_{0} = e^{i\delta}\Psi''_{1}.
\end{equation}
The equivalence class of a given section $\Psi\in S_{k}(S^{2})$ is denoted as $\tilde{\Psi}$. The representative section can now be given as
\begin{eqnarray}
\tilde{\Psi}'_{\alpha}(\chi',\chi'^{*}) &=& \sum a^{\alpha}_{n_{1}, n_{2},m_{1},m_{2}}\chi'^{*m_{1}}_{1}\chi'^{*m_{2}}_{2}\chi'^{n_{1}}_{1}\chi'^{n_{2}}_{2}~~~~ on~~~ U_{+},\\
\tilde{\Psi''}_{\alpha}(\chi'',\chi''^{*}) &=& \sum a^{\alpha}_{n_{1}, n_{2},m_{1},m_{2}}\chi''^{*m_{1}}_{1}\chi''^{*m_{2}}_{2}\chi''^{n_{1}}_{1}\chi''^{n_{2}}_{2}~~~ on~~~ U_{-},
\end{eqnarray}
with $k= m_{1}+m_{2}-n_{1}-n_{2}$. The coefficients in both the expressions are the same and the transition rule  
\begin{equation}
\tilde{\Psi}'(\chi',\chi'^{*}) = e^{ik\varphi}\tilde{\Psi}''(\chi'',\chi''^{*})
\end{equation}
is satisfied.\\
The free Dirac operator $\tilde{D}'_{k}: \tilde{S}(S^{2})\rightarrow \tilde{S}(S^{2})$ is defined by
\begin{eqnarray}
\tilde{D}'_{k} = [i\sigma'^{\mu}(\partial'_{\mu} + iA'_{\mu})] ~~~on~~~~U_{+},\\
\tilde{D}''_{k} = [i\sigma''^{\mu}(\partial''_{\mu} + iA''_{\mu})] ~~~on~~~~U_{-} ,
\end{eqnarray}
where $\partial_{\mu}$ denotes the derivative $\partial_{\theta}$, $\partial_{\varphi}$ in the local coordiantes $\theta$  and $\varphi$  in $U_{+}\bigcap U_{-}$ and ($\partial_{\mu}+iA_{\mu}$) represents covariant derivative. The $\sigma's$ ($\sigma^{\theta}$ and $\sigma^{\varphi}$) satisfies the following Clifford algebra:
\begin{equation}
\{\sigma^{\mu},\sigma^{\upsilon}\} = 2g^{\mu\nu}, ~~~\mu,\nu = 1,2  \label{rr3}
\end{equation}
in $U_{+}\bigcap U_{-}$, where$ \{g^{\mu\upsilon}\} =  \left[ {\begin{array}{cc}
1 & 0 \\
0 & 1/\sin^{2}\theta \\
\end{array} } \right]$
 is the inverse of the metric tensor \footnote{Starting from the metric  $ds^{2}  = d\chi^{\dagger}d\chi$ for $C^{2}_{0}$, we can obtain the metric on unit $S^{3}$ (r=1) by the parametrization (\ref{rr2},\ref{f1}) to get $ ds^{2} = 1/4[d\theta^{2}+d\varphi^{2}+d\psi^{2}+2\cos\theta d\varphi d\psi]$ (see appendix A3). Upon gauge-fixing $\psi = \pm\varphi$ one gets the following expressions of metric on $S^{2}$ , $ds^{2} = (d\frac{\theta}{2})^{2}+\sin^{2}\frac{\theta}{2}d\varphi^{2}$ for $U_{-}$ (\ref{c4}) and $ds^{2} = (d\frac{\theta}{2})^{2}+\cos^{2}\frac{\theta}{2}d\varphi^{2}$ for $U_{+}$ (\ref{c3}). The expression for $U_{+}$ takes the canonical form upon the replacement $\theta\rightarrow (\pi-\theta)$, as here the stereographic projection is being made from the south pole, Further since a point on $S^{2}$, having the polar angle $\theta$, subtends an angle $\theta/2$ at the south pole, it gives rise to an additional factor of 1/4 for the metric $ds^{2}$ for $S^{3}$, apart from explaining the occurrence of $\theta/2$ rather than $\theta$ in the metric for $S^{2}$. On the other hand, the metric on $S^2(\vec{x}^2=r^2)$, is of course, the canonical one: $ds^2=r^2(d\theta^2+\sin^2\theta d\phi^2)$.}
of $S^{2}$ and $A_{\mu}$ is the $k$- monopole field given by (derived in Appendix A2):
\begin{eqnarray}
A'_{\mu} &=& ik\chi'^{\dagger}\partial'_{\mu}\chi'~~~ on~~ U_{+},  \label{shu1}\\
A''_{\mu} & =& ik\chi''^{\dagger}\partial''_{\mu}\chi'' ~~~on~~~ U_{-} .\label{shu2}
\end{eqnarray}
The field $A'_{\mu}$ and $A''_{\mu}$ in $U_{+}\bigcap U_{-}$ are related by the gauge transformation
\begin{equation}
 A'_{\mu} = A''_{\mu} - ih\partial_{\mu}h^{-1} ~~,~h=e^{ik\varphi}.
\end{equation}
The $J$'s in (\ref{f4})  are the vector field on $S^{3}$ represented in the coordinate basis ($\partial_{\theta},\partial_{\varphi},\partial_{\psi}$).   These J's provide an orthonormal basis at $T_{p}(S^{3})$ and $g(J_{i},J_{j}) = \delta_{ij}$ upto an overall factor and correspond to the rotation generators in three independent directions. Correspondingly, the dual vectors $e^{i}$'s provide an orthonormal basis in the cotangent space $T^{*}_{p}(S^{3})$, $<e^{i}, J_{i}> = \delta_{ij}$. Both of these can also be obtained from the Maurer-Cartan left-invariant 1-form. Finally, we would like to mention that, using Eulerian angle parametrization, one can compute the left-invariant 1-form $e^{i}$'s occurring in the Maurer-Cartan 1-form $ g^{-1}dg$ as
 \begin{eqnarray}
  g^{-1}dg = \frac{i}{2}(e^{i}\sigma^{i})
 \end{eqnarray}
 to get
 \begin{eqnarray}
 e^{1} &=& \sin\psi d\theta -\sin\theta\cos\psi d\varphi~~~,~~~
 e^{2} = -\cos\psi d\theta - \sin\theta\sin\psi d\varphi ~~~,~~~\\
 e^{3} &=& -\cos\theta d\varphi -d\psi.
 \end{eqnarray}
They provide an orthonormal basis in $T^{*}_{p}(S^{3})$ and their duals are precisely the SO(3) rotation generators $J_{i}$ (\ref{c15},\ref{shiv}) satisfying (see appendix A3),
\begin{eqnarray}
 < e^{i}, J_{j}> &=& \delta^{i}_{j},\\
 g(J_{i},J_{j})&=& \frac{1}{4}\delta_{ij}.
\end{eqnarray} 
We should therefore be able to write them locally in the form
\begin{equation}
 e^{i} = ds^{i} ~~~, ~~~ J_{i} = \frac{\partial}{\partial s_{i}} ~~~~~i=1,2,3
\end{equation}
where the $s_{i}'s$ are the 3- affine parameters along the integral curve of $J_{i}$'s.  The tri-beins-
\begin{equation}
J_{i}\xi^{\mu} = \frac{\partial \xi^{\mu}}{\partial s_{i}} = e^{\mu}_{i} ~~~~~~\text{where},\mu=1,2,3~~ \text{s.t.}~~~~~~ \xi^{1} =\theta ~~,~~\xi^{2}=\phi~~ \text{and}~~~ \xi^{3} = \psi
\end{equation}
 and its inverse help us to relate coordinate and orthonormal basis in $T^{*}_{p}(S^{3})$ and $T_{p}(S^{3})$, respectively. Thus  by considering $\xi^{1}$ and $\xi^{2}$ only, the tri-beins help us to  connect the coordinates basis to the orthonormal basis on the cotangent $T^{*}_{p}(S^{2})$ as well.  Through the pull-back we get  
\begin{equation}
 d\xi^{\mu} = (J_{i}\xi^{\mu})ds^{i}.
\end{equation}
The $\sigma^{\mu}$ matrices, which satisfy the Clifford algebra on $S^{2}$ (\ref{rr3}), are therefore obtained from the usual Pauli matrices $\sigma^{i}$
by making use of  these tri-beins in the following way:
\begin{equation}
 \sigma^{\mu} = (J_{i}\xi^{\mu})\sigma^{i} ; ~~~\mu = 1,2.
\end{equation}
A straightforward computation yields,

  \[
  \sigma^{\theta} =
  \left[ {\begin{array}{cc}
   1 & -\cot\theta e^{-i\varphi} \\
   -\cot\theta e^{i\varphi} & -1 \\
  \end{array} } \right],
\]
 \[
  \sigma^{\varphi} =
  \left[ {\begin{array}{cc}
   0 & -i e^{-i\varphi} \\
   i e^{i\varphi} & 0 \\
  \end{array} } \right],
\]
 satisfying (\ref{rr3}) and the corresponding expressions for the connection components (\ref{shu1},\ref{shu2}) yield
\begin{eqnarray}
 A'_{\theta} = 0 ,~~~~A'_{\varphi} = \frac{k}{2}(\cos\theta-1),\\
 A''_{\theta} = 0 ,~~~~A''_{\varphi} = \frac{k}{2}(\cos\theta+1).
\end{eqnarray}
In $S^{2}$ the  eigenvalue problem of the Dirac operator has to do in the patches $U_{+}$ and $U_{-}$ differently, so the $\tilde{D}_{k}$ in $S_{k}(S^{2})$ is switched to the problem in $S_{k}(S^{3})$, which will be defined globally on $S^{3}$, with the transformation of the sections and $\tilde{D}_{k}$
\begin{eqnarray}
 \Psi = e^{-\frac{i}{2}k(\varphi+\psi)}\tilde{\Psi}' ~~~on~~ U_{+},\\
 \Psi = e^{\frac{i}{2}k(\varphi-\psi)}\tilde{\Psi}' ~~~on~~ U_{-},
 \end{eqnarray}
 and 
 \begin{eqnarray}
 D_{k} = e^{-\frac{i}{2}k(\varphi+\psi)}\tilde{D}_{k}e^{\frac{i}{2}k(\varphi+\psi)} ~~on ~~U_{+},\\
 D_{k} = e^{-\frac{i}{2}k(\varphi+\psi)}\tilde{D}_{k}e^{\frac{i}{2}k(\varphi+\psi)}~~on~~U_{-}.
\end{eqnarray}
A straightforward computation shows that these are special cases of the Dirac operator on $S^{3}$ given by
\begin{equation}
 D_{k} = \frac{1}{r}\sigma_{j}(J_{j}-\frac{k}{2}\frac{x_{j}}{r}). \label{c30}
\end{equation}

\section{Spectral triplet on the configuration space}

In the commutative case the complex valued functions defined on the manifold serve the purpose of the elements of the algebra $(\mathcal{A})$.  The function of $\chi$'s , $\chi^{\dagger} s\in S^{3}$ of the type -
\begin{equation}
\Psi = \sum_{m_{1},m_{2},n_{1},n_{2}}\chi_{1}^{*m_{1}}\chi^{*^{m_{2}}}_{2}\chi^{n_{1}}_{1}\chi^{n_{2}}_{2} \label{c17}
\end{equation}
with $k= m_{1}+m_{2}-n_{1}-n_{2} = 0$ are the  functions defined on $S^{2}$, which is clear from (\ref{f10}), since the $\psi$ dependence disapears . The composition of any two such functions (\ref{c17}) preserve the condition $k=0$, so that they indeed form an algebra. Thus, in the commutative case, we have the algebra $(\mathcal{A})$:
\begin{equation}
\mathcal{A} = \{\Psi = \sum\chi_{1}^{*m_{1}}\chi^{*^{m_{2}}}_{2}\chi^{n_{1}}_{1}\chi^{n_{2}}_{2} ~~\rvert~~ k= m_{1}+m_{2}-n_{1}-n_{2} = 0 \}. \label{bh10}
\end{equation}

In the non-commutative case we can also take the following algebra with elements as in (\ref{c17}) but the $\chi's$ and $\chi^{\dagger}$'s are now operators, which are taken in the normal-ordered form:
\begin{equation}
\mathcal{A} = \{\hat{\Psi} = \hat{\chi}_{1}^{\dagger m_{1}}\hat{\chi}^{\dagger^{m_{2}}}_{2}\hat{\chi}^{n_{1}}_{1}\hat{\chi}^{n_{2}}_{2} ~~\rvert~~ k= m_{1}+m_{2}-n_{1}-n_{2} = 0 \}. \label{c25}
\end{equation}

The previously defined dilatation operator $\hat{K}$ and rotation generators $\hat{J}_{i}$ (\ref{c15}, \ref{c16}) in the commutative case  acts on the section of the  spinor bundle. The action of $\hat{J}_{i}$ and $\hat{K}$ on the $\Psi_{\alpha}$ in the non-commutative case is given by the following adjoint actions:
\begin{eqnarray}
\hat{J}_{i}\Psi_{\alpha} &=& \frac{1}{\lambda}[x_{i},\Psi_{\alpha}], \label{g3}\\
\hat{K}\Psi_{\alpha} &=& [\hat{N},\Psi_{\alpha}].  \label{c22} ~~~~~,~~~~~ 
\end{eqnarray}
These actions of $\hat{J}_{i}$ and $\hat{K}$ in the non-commutative case is obtained by replacing the derivatives in (\ref{c15}, \ref{c16}) with the following  commutators:
\begin{eqnarray}
\partial_{\chi_{\alpha}}\Psi \sim [\chi^{\dagger}_{\alpha},\Psi]  ~~,~~\partial_{\chi^{\dagger}_{\alpha}}\Psi \sim [\chi_{\alpha},\Psi].
\end{eqnarray}
Since $\hat{J}_{i}$ acts on $\Psi_{\alpha}\in \mathcal{H}_n$ adjointly, these operators will act on $\mathcal{F}_c$ simply as
\begin{equation}
  \hat{J}_i|n,n_3\rangle = \frac{1}{\lambda}\hat{X}_i|n,n_3\rangle. \label{j3action}
 \end{equation}

One can easily see that those operators for which $k=0$ are those acting on $\mathcal{F}_{n}$ (\ref{c20}).  Since the operator $\hat{K}$ (\ref{c22})  has the eigenvalue $k=0$ on these operators $\hat{\Omega}= \rvert n,n_{3}\rangle\langle n,n'_{3}\rvert $, i.e.
\begin{equation}
K \rvert n,n_{3}\rangle\langle n,n'_{3}\rvert = [\hat{N} ,\rvert n,n_{3}\rangle\langle n,n'_{3}\rvert] = 0. \label{g1}
\end{equation}
These $\hat{\Omega}$ should therefore be the analogues of the $\Psi$ (\ref{c17}) in the non-commutative case.  In the non-commutative case we therefore take the algebra $\mathcal{A}$ to be
\begin{equation}
\mathcal{A} = Span\{\rvert n,n_{3}\rangle\langle n,n'_{3}\rvert~~ \rvert~~ -n\leq n_{3},n'_{3}\leq n\}. \label{c26}
\end{equation}
 The Hilbert space ($\mathcal{H}$) on which the algebra (through appropriate representation $\pi(a)=\begin{pmatrix}
                                                                                                     a&0\\
                                                                                                     0&a
                                                                                                    \end{pmatrix}$
with $a\in \mathcal{A}$) and Dirac operator act is taken to be the subspace of the classical configuration space $\mathcal{F}_n$ tensored with $\mathbb{C}^2$. The Dirac operator $(\mathcal{D})$ is given in (\ref{c30}) with, of course, the $J_{i}$ 's occurring there now corresponding to operators with their actions on $\mathcal{F}_n$ given in (\ref{j3action}).
 
 Let us now write the spectral triple for this fuzzy sphere.
\begin{enumerate}
\item {The Algebra $\mathcal{A}$ = Span\{$\rvert n,n_{3}\rangle\langle n,n'_{3}\rvert , -n\leq n_{3},n'_{3}\leq n$\}}.
\item{The Hilbert space $\mathcal{H}$ =  $\mathcal{F}_n\otimes \mathbb{C}^2 = \Big\{\left( 
\begin{array}{c}
|n,n_3\rangle\\
|n,n'_3\rangle
\end{array}
\right)\Big\} $ }.
\item{Dirac operator $\mathcal{D} = D_{k} = \frac{1}{r}\sigma_{j}(\hat{J}_{j}-\frac{k}{2}\frac{\hat{x}_{j}}{r}) $}.
\end{enumerate}
Note that the Dirac operator is taken to be of the same form as in (\ref{c30}),and that $\big(\frac{1}{r}\big)$ occurring in the front is taken to be the constant $\big(\frac{1}{\lambda\sqrt{n(n+1)}}\big)$ following from (\ref{r^2no.})-appropriate for the reciprocal of the radius of the $n$th sphere.

Like in the Moyal case \cite{b1}, the "points" of the non-commutative space are given by the pure states of the algebra which belong to the dual of the algebra (here algebra and its dual are same ). The pure state density matrices $\rvert n,n_{3}\rangle \langle n,n_{3}\rvert$ corresponding to $\rvert n,n_{3}\rangle\in \mathcal{F}_n$ where $-n\leq n_{3}\leq n$ represent  the "points" of the fuzzy/noncommutative sphere whose radius is indexed by n.

\subsection{Connes distance calculation on the fuzzy sphere}
 
A fuzzy sphere  is  described by the Hilbert space $\mathcal{F}_{n}$. To give the notion of distance on such fuzzy space, we will use the Connes spectral distance formula (\ref{sp.dist.}).

We would like to calculate the distance between the two nearest generalized points on the fuzzy sphere which we can obtain by  applying the distance formula (\ref{sp.dist.}) between the pure states of the algebra $\mathcal{A}$, which are in fact the pure density matrices, given by $\rvert n ,n_{3}\rangle\langle n,n_{3}\rvert$ and $\rvert n ,n_{3}+1\rangle\langle n,n_{3}+1\rvert$ acting on the Hilbert space $\mathcal{F}$. 
The compact form of  the distance formula between the pure density matrices $\rho$ and $\rho'$, which was given in  \cite{b1} and is adaptable for the Hilbert-Schmidt operatorial formulation, is in this case given by 
\begin{equation}
d(\rho,\rho') = Sup\{\rvert a(\rho)-a(\rho')\rvert ~~~~,~~\lVert[\mathcal{D}, a]\rVert_{op}\leq 1, a\in\mathcal{A}\} = \frac{\text{tr}(d\rho^{2})}{\lVert[D,\pi(d\rho)]\rVert_{op}}.  \label{opdis}
\end{equation}
Here $\rho  = \rvert n,n_{3}\rangle\langle n,n_{3}\rvert$ and $\rho' = \rvert n,n_{3}+1\rangle\langle n,n_{3}+1\rvert$ and $d\rho$ is
\begin{equation}
d\rho =  \rho'-\rho  = \rvert n,n_{3}+1\rangle\langle n,n_{3}+1\rvert - \rvert n,n_{3}\rangle\langle n,n_{3}\rvert.
\end{equation}
First we need to calculate $\lVert[\mathcal{D},\pi(d\rho)]\rVert_{op}$, but before we proceed further, we note that $d\rho$, by itself, is a Hilbert-Schmidt operator with $k=0$ and therefore belongs to the algebra $\mathcal{A}$. Consequently, we can set $k=0$ in $\mathcal{D}$ (\ref{c30}) right in the beginning, so that $\mathcal{D}$ effectively takes the following form:
\begin{eqnarray}
 \mathcal{D} = \frac{1}{r}\left[ {\begin{array}{cc}
J_{3} & J_{1}-iJ_{2}\\
J_{1}+iJ_{2} & -J_{3} \\
\end{array} } \right] .\label{g2}
\end{eqnarray}
Since the action of $J_i$, as given in (\ref{j3action}), on the Hilbert is given by the left action, the commutator  $[\mathcal{D},\pi(d\rho)]$ reduces to 
\begin{eqnarray}
[\mathcal{D},\pi(d\rho)] &=&  \frac{1}{r}\left[ {\begin{array}{cc}
0 & \frac{1}{\lambda}[(\hat{x}_{1}-i\hat{x}_{2}),d\rho]\\
\frac{1}{\lambda}[(\hat{x}_{1}+i\hat{x}_{2}),d\rho] & 0 \\
\end{array} } \right] \\
&=& \frac{1}{r} \left[ {\begin{array}{cc}
0 & \frac{1}{\lambda}[\hat{x}_{-},d\rho]\\
\frac{1}{\lambda}[\hat{x}_{+},d\rho] & 0 \\
\end{array} } \right],
\end{eqnarray}
where the commutator $[\hat{x}_{3},d\rho]$ vanishes. In order to take the norm we compute
\begin{eqnarray}
[\mathcal{D},\pi(d\rho)]^{\dagger}[\mathcal{D},\pi(d\rho)] = \frac{1}{r^{2}}\left[ {\begin{array}{cc}
 \frac{1}{\lambda^{2}}[\hat{x}_{-},d\rho]^{\dagger}[\hat{x}_{-},d\rho]& 0 \\
0 &  \frac{1}{\lambda^{2}}[\hat{x}_{+},d\rho]^{\dagger}[\hat{x}_{+},d\rho]\\
\end{array} } \right].
\end{eqnarray}
An explicit computation yields
\begin{eqnarray}
\frac{1}{\lambda}[\hat{x}_{+},d\rho] = \sqrt{n(n+1)-(n_{3}+1)(n_{3}+2)}\rvert n,n_{3}+2\rangle\langle n,n_{3}+1\rvert \nonumber\\-2\sqrt{n(n+1)-n_{3}(n_{3}+1)}\rvert n,n_{3}+1\rangle\langle n,n_{3}\rvert \nonumber \\+\sqrt{n(n+1)-n_{3}(n_{3}-1)}\rvert n,n_{3}\rangle\langle n,n_{3}-1\rvert,
\end{eqnarray}
\begin{eqnarray}
\frac{1}{\lambda}[\hat{x}_{-},d\rho] = \sqrt{n(n+1)-(n_{3}+1)(n_{3}+2)}\rvert n,n_{3}+1\rangle\langle n,n_{3}+2\rvert \nonumber \\+2\sqrt{n(n+1)-n_{3}(n_{3}+1)}\rvert n,n_{3}\rangle\langle n,n_{3}
+1\rvert \nonumber\\+\sqrt{n(n+1)-n_{3}(n_{3}-1)}\rvert n,n_{3}-1\rangle\langle n,n_{3}\rvert, 
\end{eqnarray}
so that 
\begin{eqnarray}
\frac{1}{\lambda^2}[\hat{x}_{+},d\rho]^{\dagger}[\hat{x}_{+},d\rho] = [n(n+1)-(n_{3}+1)(n_{3}+2)]\rvert n,n_{3}+1\rangle\langle n,n_{3}+1\rvert \nonumber\\ +4[n(n+1)-(n_{3}+1)n_{3}]\rvert n,n_{3}\rangle\langle n,n_{3}\rvert \nonumber\\+[n(n+1)-(n_{3}-1)n_{3}]\rvert n,n_{3}-1\rangle\langle n,n_{3}-1\rvert, 
\end{eqnarray}
\begin{eqnarray}
\frac{1}{\lambda^2}[\hat{x}_{-},d\rho]^{\dagger}[\hat{x}_{-},d\rho] = [n(n+1)-(n_{3}+1)(n_{3}+2)]\rvert n,n_{3}+2\rangle\langle n,n_{3}+2\rvert \nonumber\\ +4[n(n+1)-(n_{3}+1)n_{3}]\rvert n,n_{3}+1\rangle\langle n,n_{3}+1\rvert \nonumber\\+[n(n+1)-(n_{3}-1)n_{3}]\rvert n,n_{3}\rangle\langle n,n_{3}\rvert.
\end{eqnarray}
 Since both of these operators are diagonal, the operator norm,  defined to be the largest eigenvalue,  can be read off exactly from both these equations, yielding $4[n(n+1)-n_{3}(n_{3}+1)]$. This gives the operator norm of the commutator as,
\begin{equation}
\lVert[\mathcal{D},\pi(d\rho)]\rVert  =\frac{2\sqrt{[n(n+1)-n_{3}(n_{3}+1)]}}{\lambda\sqrt{n(n+1)}}.
\end{equation}
The infinitesimal distance on the fuzzy sphere is then easily obtained from (\ref{opdis}) as 
\begin{equation}
d(n_{3}+1,n_{3}) = \frac{\lambda\sqrt{n(n+1)}}{\sqrt{[n(n+1)-n_{3}(n_{3}+1)]}},    \label{rr1}
\end{equation}
where we have used $\text{tr}_{c}(d\rho)^{2} =2$.

\subsection{On the infinitesimal nature of the distance formula (\ref{rr1})}


To clarify the adjective "infinitesimal" in this context, let us recall from the theory of angular momentum that the state $\rvert n,n_{3} \rangle$ can be visualized as the vector $\vec{x}$ precessing the $x_{3}$- axis along a cone, in such a manner that the tip of the vector $\vec{x}$ lies on the circle of latitude on a sphere of radius $\lambda\sqrt{n(n+1)}$, maintaining a fixed $x_{3}$- component $\lambda n_{3}$ with $n_{3}$ varying in the interval $-n\leq n_{3}\leq n (n\in \mathbb{Z}/2)$ in the steps of unity  (see fig.1). The associated polar angles are therefore quantized as,
\begin{eqnarray}
\theta_{n_{3}} = \sin^{-1}(\frac{n_{3}}{\sqrt{n(n+1)}}). \label{bh1}
\end{eqnarray}
Now, let us treat $n_{3}$ to be a continuous variable for a moment. This yields
\begin{eqnarray}
 d\theta_{n_{3}} = \frac{dn_{3}}{\sqrt{n(n+1)-n_{3}^{2}}}.
\end{eqnarray}
The distance, which is identified with arc length in figure 1, is then obtained by multiplying with the quantized radius to get
\begin{eqnarray}
 ds(n_{3}) = \frac{\lambda\sqrt{n(n+1)}dn_{3}}{\sqrt{n(n+1)-n_{3}^{2}}} .\label{rup1}
\end{eqnarray}
This almost matches with the distance expression (\ref{rr1}); in fact with the formal replacement $dn_{3}\rightarrow \Delta n_{3} =1 ~~\text{and}~~ n_{3}^{2}\rightarrow n_{3}(n_{3}+1)$ one reproduces (\ref{rr1}) exactly. Further, (\ref{rup1}) can be shown to follow, albeit somewhat heuristically, from (\ref{rr1}) by taking the average of the $n_{3}$ dependence in (\ref{rr1}), which gives $d(n_{3},n_{3}+1)$  and that of the $n_{3}$ -dependence occurring in $d(n_{3}-1,n_{3}):  ~\frac{1}{2}[n_3(n_3+1)+n_3(n_3-1)]=n_3^2$. It is noted that the same trick works in the Moyal case also \cite{b1}.

\begin{tikzpicture} 

\def\R{2.5} 
\def\angEl{15} 
\def\angAz{-105} 
\def\angPhi{-40} 
\def\angBeta{19} 
\draw[help lines, thin] (0,0) circle (\R);
\foreach \t in {-80,-60,...,80} { \DrawLatitudeCircle[\R]{\t} }
\foreach \t in {40,60} { \DrawLatitudeCircleL[\R]{\t} }
\foreach \t in {-45} { \DrawLongitudeCircle[\R]{\t} }

\pgfmathsetmacro\H{\R*cos(\angEl)}
\coordinate[mark coordinate] (N) at (0,\H) ;
\coordinate[mark coordinate] (S) at (0,-\H);
\coordinate (O) at (0,0,0);
\node[right] at (0,2.8,0) {N};
\node[right] at (0,-2.8,0) {S};
\draw [->] (.5,.45) to [out=60,in=-15] (.3,.6);
\node [above right] at (.35,.45) {\small$d\theta_{n_3}$}; 
\draw[help lines,thick,->] (0,0,0) -- (4,0,0);
\node[right] at (4,0,0) {$\hat{x}_2$};
\draw[help lines,thick,->] (0,0,0) -- (0,4,0);
\node[above] at (0,4,0) {$\hat{x}_3$};
\draw[help lines,thick,->] (0,0,0) -- (0,0,6);
\node[left] at (0,0,6) {$\hat{x}_1$};
\draw[->] (0,0,0) -- (2.7,4,4);
\draw[->] (0,0,0) -- (4.35,4,4);
\draw[help lines,<-] (0,1.85,0) -- (2.45,3.41,4);
\draw[help lines,<-] (0,1.2,0) -- (2.15,2,2.05);
\draw[thin,decorate,decoration={brace,raise=0.5pt,amplitude=1ex}] (0,0,0) -- (0,1.2,0)
    node[midway,left=1ex] {\small $\lambda n_3$};
\draw[thin,decorate,decoration={brace,raise=0.5pt,amplitude=2ex}] (0,0,0) -- (0,1.85,0)
    node[midway,left=2ex] {\small $\lambda (n_3+1)$};   
\draw[thin,decorate,decoration={brace,raise=0.5pt,amplitude=0.5 ex}] (0,1.2,0) -- (0,1.85,0)
    node[midway,left=0.5 ex] {\small $\lambda \Delta n_3=\lambda$~~};
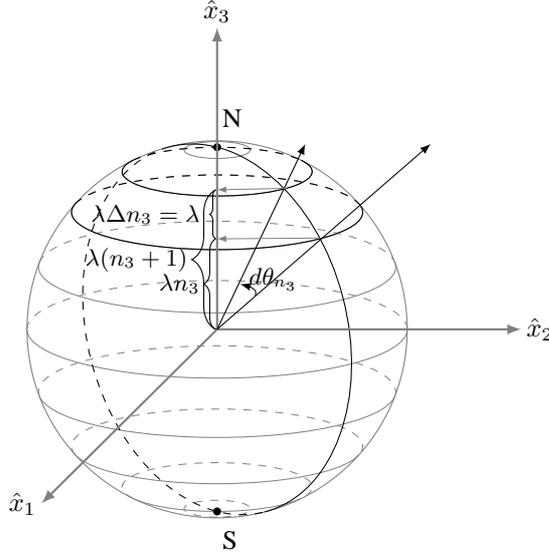
\node[below] at (0,-4,0) {Fig. 1: Visualization of infinitesimal change on the surface of sphere with respect to the change in $n_3$};
\end{tikzpicture}

\subsection{Construction of Perelemov Coherent state on $S^{2}$ and the Connes distance function}

 We now provide a brief review of the construction of Perelemov's $SU(2)$ coherent state, as given in \cite{b5}. Let us consider a general Lie group $G$, whose unitary irreducible representation on some Hilbert space $\mathcal{H}$ is denoted as $T(g)$. Consider a fixed vector in the Hilbert space denoted as $\rvert x_{0}\rangle$ and consider $\rvert x\rangle$ obtained as, $\rvert x\rangle=T(g)\rvert x_{0}\rangle$, where $g$ is any element of the group $G$. The two state $T(g_{1})\rvert x_{0}\rangle$ and $T(g_{2})\rvert x_{0}\rangle$ are called equivalent if they differ by a phase factor-
\begin{eqnarray}
T(g_{1})\rvert x_0\rangle = e^{i\alpha}T(g_{2})\rvert x_{0}\rangle \Rightarrow T(g^{-1}_{2}g_{1})\rvert x_{0}\rangle = e^{i\alpha}\rvert x_{0}\rangle ~~,~~~\rvert e^{i\alpha}\rvert = 1.
\end{eqnarray}
Consider the subgroup $H$ of the group $G$ with the property
\begin{equation}
T(h)\rvert x_{0}\rangle = e^{i\alpha(h)}\rvert x_{0}\rangle.
\end{equation}
This construction shows that the vectors $\rvert x\rangle_{g}$ for all group element g, which belong to the same equivalence class, determined by the left coset class of group $G$ with respect to the subgroup $H$, differs only in the phase factor, so they determine the same state. Choosing a representative  $g(x)$ in the equivalence class $x\in G/H$ ,one gets the associated state $\rvert x\rangle$. These states are the Perelemov coherent states on the base manifold $G/H$. Here we are particularly interested in the construction of coherent states on $S^{2}= SU(2)/U(1)$, with $G= SU(2)$ and $H=U(1)$. We write an element of $SU(2)$ as the matrix
\begin{eqnarray}
 g=\left[ {\begin{array}{cc}
u & v\\
-\bar{v} & \bar{u} \\
\end{array} } \right]~;~~~|u|^2+|v|^2=1,
\end{eqnarray}
which can be parametrized in terms of Euler angles exactly like (\ref{f1},\ref{rr2}).
The subgroup $H=U(1)$
\begin{eqnarray}
 h=\left[ {\begin{array}{cc}
u & 0\\
0 & \bar{u} \\   
\end{array} } \right] ~~~~~,~~~~~ u=e^{\frac{-i}{2}\psi}
\end{eqnarray}
is the stability subgroup. As discussed earlier the $CP^{1}$ =  $S^{2}$ manifold is described by the doublet $\left(
\begin{array}{c}
1\\
z\\
\end{array}
\right)$. Upon normalization, one can identify, $u= 1/\sqrt{(1+\rvert z\lvert^{2})}$ and $v= z/\sqrt{(1+\lvert z\rvert^{2})}$ , so that the associated  $SU(2)$ group element can be written as 
\begin{eqnarray}
 g = \frac{1}{\sqrt{1+\lvert z\rvert^{2}}}\left[ {\begin{array}{cc}
1 & z\\
-\bar{z} & 1 \\
\end{array} } \right].   \label{i1}
\end{eqnarray} 
Here $z\in C$ represents the stereographic projected coordinates of the points of the $S^{2}$ manifold from the south pole. The coherent state on the $S^{2}$ is then 
\begin{equation}
 \rvert z\rangle =T(g(z))\rvert 0\rangle,  \label{saur1}
\end{equation}
where $\rvert x_{0}\rangle\equiv\rvert 0\rangle$ represents the north pole, and $T(g(z)$) is given by (\ref{i1}). This can be recast as
\begin{eqnarray}
\rvert z\rangle = \frac{1}{\sqrt{1+\lvert z\rvert^{2}}}\rvert 0\rangle +\frac{1}{\sqrt{1+\lvert z\rvert^{2}}}(z\sigma_{+}-\bar{z}\sigma_{-})\rvert 0\rangle,
\end{eqnarray}
where $\sigma_{\pm} = \sigma_{1}\pm i\sigma_{2}$ and $\sigma_{1} ,\sigma_{2}$ are the two Pauli matrices in the  fundamental $j=1/2$  representation of the $su(2)$ Lie algebra. In an arbitrary  $j$-th representation,  we can write the coherent state as-
\begin{eqnarray}
\rvert z\rangle = \frac{1}{\sqrt{1+\lvert z\rvert^{2}}}\rvert 0\rangle +\frac{1}{\sqrt{1+\lvert z\rvert^{2}}}(z J_{+}-\bar{z} J_{-})\rvert 0\rangle ~~,~~J_{\pm} = J_{1}\pm iJ_{2} 
\end{eqnarray}
 with $J_{i}$ being the corresponding generators. In the non-commutative case the $\rvert 0\rangle$, which represents the north pole, is identified with the state $\rvert n,n_{3}=n\rangle (r=\lambda\sqrt{n(n+1)}, x_{3} = \lambda n)$, which represent the north pole closely.  We can therefore write the coherent state as
\begin{eqnarray}
\rvert z\rangle = \frac{1}{\sqrt{1+\lvert z\rvert^{2}}}\rvert n,n\rangle +\frac{1}{\sqrt{1+\lvert z\rvert^{2}}}(z J_{+}-\bar{z} J_{-})\rvert n,n\rangle.
\end{eqnarray}
In order to compute infinitesimal distance function on $S^{2}$ using the coherent state, we write  $d\rho$ as in the Moyal case \cite{b1} as
\begin{eqnarray}
d\rho = \rvert z+dz\rangle\langle z+dz\rvert -\rvert z\rangle\langle z\rvert. \label{ark1}
\end{eqnarray}
 Now the left invariant Maurer-Cartan 1-form  $g^{-1}(z)dg(z)$ can be easily computed, using the SU(2) parametrization (\ref{i1}) to get (in the $j=1/2$ representation)
\begin{eqnarray}
 g^{-1}(z)dg(z) = i\frac{A}{2}\sigma_{3} +\frac{i}{1+\rvert z\rvert^{2}}[-i\sigma_{+}dz+i\sigma_{-}d\bar{z}], \label{fu1}
\end{eqnarray}
 where 
 \begin{eqnarray}
  A = i(\frac{\bar{z}dz-zd\bar{z}}{1+\lvert z\rvert^{2}})
 \end{eqnarray}
is the U(1) connection 1-form derived in (\ref{io1}) of Appendix A2. The state $\rvert z+dz\rangle$ is written as
\begin{eqnarray}
 \rvert z+dz\rangle = (1+g^{-1}dg-i\frac{A}{2}\sigma_{3})\rvert z\rangle. \label{pp4}
\end{eqnarray}
 Note that we have excluded the $\sigma_{3}$ term, as this is associated  with the stability subgroup $U(1)$. To explain the strategy we have adopted here in a better way, we observe that any point on $S^{2}$ can be regarded as a north pole, upon suitable $SO(3)$ rotation of the coordinate axes and $J_{i}'s $ can be regarded as the $SU(2)$ generators, associated to these new rotated axes.  With this the point coordinatized by $z$, will remain invariant under the action of the stability subgroup $U(1)$ contained in $SU(2)$, for which the generator is $J_{3}$. Consequently, the state $\rvert z\rangle$ will now correspond to $\rvert n,n\rangle$ and $J_{3}$ terms or rather $\sigma_{3}$ (\ref{fu1}) can be disregarded.  This finally yields in the $j=1/2$ representation
\begin{eqnarray}
\rvert z+dz\rangle =  \rvert z\rangle +\frac{i}{1+\rvert z\rvert^{2}}[-i\sigma_{+}dz+i\sigma_{-}d\bar{z}]\rvert z\rangle.
\end{eqnarray}
More accurately, we should replace $\frac{\sigma_{\pm}}{2}\rightarrow J_{\pm}$ to describe coherent state on a generic sphere of radius $\lambda\sqrt{n(n+1)}$, as it will now correspond to the $j=n$ representation of $(2n+1)$ dimension. Correspondingly, we should write 
\begin{eqnarray}
 \rvert z+dz\rangle = \rvert z\rangle +\frac{2i}{1+\lvert z\rvert^{2}}[-iJ_{+}dz+iJ_{-}d\bar{z}]\rvert z\rangle. \label{asu1}
\end{eqnarray}
Since $\rvert z\rangle$ is identified with $\rvert z\rangle \equiv \rvert n,n\rangle$, by rotating the coordinate axes, we can simplify (\ref{asu1}) to get
\begin{eqnarray}
 \rvert z+dz\rangle = \rvert z\rangle +\frac{i\sqrt{2n}d\bar{z}}{1+\rvert z\rvert^{2}}\rvert n,n-1\rangle.
\end{eqnarray}
A straightforward computation gives, using (\ref{ark1}),
\begin{eqnarray}
 d\rho = \frac{i\sqrt{2n}}{1+\rvert z\rvert^{2}}(d\bar{z}\rvert n,n-1\rangle\langle n,n\rvert -dz\rvert n,n\rangle\langle n,n-1\rvert) \label{drho}.
\end{eqnarray}

With the same spectral triple as above, we can define the infinitesimal distance on the fuzzy sphere for the continuous case taking the  $d\rho$ (\ref{drho}) above. We have the same Dirac operator and so the same commutator formula:
\begin{eqnarray}
 [\mathcal{D},\pi(d\rho)] = \frac{1}{r} \left[ {\begin{array}{cc}
 0 & \frac{1}{\lambda}[x_{-}, d\rho]\\
\frac{1}{\lambda}[x_{+},d\rho] & 0 \\   
\end{array} } \right]
\end{eqnarray}
and
\begin{eqnarray}
 [\mathcal{D},\pi(d\rho)]^{\dagger} [\mathcal{D},\pi(d\rho)] = \frac{1}{r^{2}} \left[ {\begin{array}{cc}
 \frac{1}{\lambda^{2}}[x_{-}, d\rho]^{\dagger} [x_{-}, d\rho] & 0 \\
 0 &  \frac{1}{\lambda^{2}}[x_{+},d\rho]^{\dagger} [x_{+}, d\rho]\\   
\end{array} } \right].
\end{eqnarray}
After computation, we get
\begin{eqnarray}
\frac{1}{\lambda}[x_{+}, d\rho] = \frac{i\sqrt{2n}}{1+\lvert z\rvert^{2}}(d\bar{z}\sqrt{2n}\rvert n\rangle\langle n\rvert - d\bar{z}\sqrt{2n}\rvert n-1\rangle\langle n-1\lvert \nonumber\\+dz\sqrt{2(2n-1)}\rvert n\rangle\langle n-2\rvert),
\end{eqnarray}
\begin{eqnarray}
\frac{1}{\lambda}[x_{-}, d\rho] = \frac{i\sqrt{2n}}{1+\lvert z\rvert^{2}}(d\bar{z}\sqrt{2(2n-1)}\rvert n-2\rangle\langle n\rvert -dz\sqrt{2n}\rvert n-1\rangle\langle n-1\lvert \nonumber \\+dz\sqrt{2n}\rvert n\rangle\langle n\rvert).
\end{eqnarray}
Here we have kept in mind that it is $\frac{x_{\pm}}{\lambda}$, rather than $x_{\pm}$ that behave as ladder operators.
The maximum eigenvalue $\lambda_{max}$ is obtained by diagonalization to yield,
\begin{eqnarray}
 \lVert[\mathcal{D}, \pi(d\rho)]\rVert_{op} = \sqrt{\frac{4n(3n-1)}{(1+\lvert z\rvert^{2})^{2}}d\bar{z}dz}.
\end{eqnarray}
Finally, using
\begin{eqnarray}
 \text{tr}_{c}(d\rho)^{2} = \frac{4n}{(1+\lvert z\rvert^{2})^{2}}d\bar{z}dz,
\end{eqnarray}
 and multiplying by the radius $\lambda\sqrt{n(n+1)}$  we get the infinitesimal distance on $S^{2}$: 
\begin{eqnarray}
 d(\lvert z\rangle, \rvert z+dz\rangle) = \frac{\lambda\sqrt{n(n+1)}}{1+\rvert z\rvert^{2}}\sqrt{\frac{4n}{3n-1}d\bar{z}dz} =\lambda\sqrt{\frac{4n^{2}(n+1)}{3n-1}}\frac{\sqrt{d\bar{z}dz}}{1+\rvert z\rvert^{2}}. \label{pp1}
\end{eqnarray}
This is precisely, the form of the metric on $S^{2}$ in the stereographic variable. Further the linear scaling with $n$ for large $n$ is manifest in the expression.

\section{Spectral Distance on the quantum Hilbert space}

The quantum Hilbert space is spanned by the Hilbert-Schmidt operators acting on the fuzzy sphere described by $\mathcal{F}_n$. That is,
\begin{equation}
 \mathcal{H}_n= Span\{\rvert n,n_{3}\rangle\langle n,n'_{3}\rvert\}\equiv |n_3,n'_3):~~\text{tr}_c(\Psi^\dag\Psi)<\infty\}.
\end{equation}
Since we consider a particular fuzzy sphere indexed by $n$, we have suppressed the index $n$ above and by taking analogy with the case of Moyal plane, we can construct the spectral triple for this case as:
 
\begin{enumerate}
\item {The Algebra $\mathcal{A} = Span\{|n_3,n'_3)(l_3,l'_3| :~~ -n\leq n_3,n'_3,l_3,l'_3\leq n$ ~ with ~$n$~ being fixed\}}.
\item{The Hilbert space $\mathcal{H}$ =  $\mathcal{H}_n\otimes \mathbb{C}^2 = \Big\{\left( 
\begin{array}{c}
|n_3,n'_3)\\
|l_3,l'_3)
\end{array}
\right) \Big\}$ }. 
\item{Dirac operator $\mathcal{D} =  \frac{1}{r}\sigma_{j}\hat{J}_{j} $ },
\end{enumerate}
where we have taken $k=0$ and the action of $J_i$ on $\mathcal{H}_q$ is given in (\ref{g3}).

In this case, we can define spectral distance between both pure and mixed states of the algebra. First consider the pure states of the algebra $\mathcal{A}$ corresponding to the density matrices $\rho_q(n_3, n'_3) = |n_3,n'_3)(n_3,n'_3|$  and $\rho_q(n_3+1, l'_3) = |n_3+1,l'_3)(n_3+1,l'_3|$. Then we have the operator $d\rho_q = |n_3+1,l'_3)(n_3+1,l'_3|- |n_3,n'_3)(n_3,n'_3|$, which should reproduce the infinitesimal distance between states computed earlier in (\ref{rr1}) when we take $n'_3=l'_3$. To this end, let us begin by computing 
\begin{equation}
[\mathcal{D},\pi(d\rho_q)] = \frac{1}{r} \begin{bmatrix}
0 & \frac{1}{\lambda}[\hat{X}_{-},d\rho_q] \\
\frac{1}{\lambda}[\hat{X}_{+},d\rho_q] & 0 
\end{bmatrix},  
\end{equation}
so that
\begin{equation}
[\mathcal{D},\pi(d\rho_q)]^{\ddagger}[\mathcal{D},\pi(d\rho_q)] = \frac{1}{r^2}  \begin{bmatrix}
 \frac{1}{\lambda^2}[\hat{X}_{-},d\rho_q]^{\ddagger}[\hat{X}_{-},d\rho_q]& 0 \\
0 &  \frac{1}{\lambda^2}[\hat{X}_{+},d\rho_q]^{\ddagger}[\hat{X}_{+},d\rho_q]\\
\end{bmatrix} .
\end{equation}
Here, $\hat{X}_i$ and $\hat{X}_{\pm}$ are the position operators and the corresponding ladder operators acting on $\mathcal{H}_q$.

After computation, we get 
 \begin{equation}
  \lVert [\mathcal{D},\pi(d\rho_q)]\rVert_{\text{op}}=\begin{cases}
    \frac{2\sqrt{[n(n+1)-n_{3}(n_{3}+1)]}}{\lambda\sqrt{n(n+1)}}, & \text{if}~~ n'_3=l'_3.\\
    \frac{\sqrt{[n(n+1)-n_{3}^2 +|n_{3}|]}}{\lambda\sqrt{n(n+1)}}, & \text{otherwise}.
  \end{cases}
\end{equation}
Since here we also have tr$_q (d\rho_q)^2=2$, we get the infinitesimal distance on the quantum Hilbert space by using a formula with the same form as that of (\ref{opdis}): 
\begin{equation}
 d(\rho_q(n_3+1, l'_3),\rho_q(n_3, n'_3) )=\begin{cases}
                                            \frac{\lambda\sqrt{n(n+1)}}{\sqrt{[n(n+1)-n_{3}(n_{3}+1)]}}, & \text{if}~~ n'_3=l'_3.  \\
                                            \frac{2\lambda\sqrt{n(n+1)}}{\sqrt{[n(n+1)-n_{3}^2 +|n_{3}|]}}, & \text{otherwise}.
                                           \end{cases}
\end{equation}
This shows that just like in the Moyal case \cite{b1} the distance on quantum Hilbert space $\mathcal{H}_n$ of the fuzzy sphere depends on the right hand sectors and it increases when the right hand sectors are taken differently, although the Dirac operator acts only on the left hand sector.

Now let us consider a more general situation where the density matrices is of the mixed form and is given by
\begin{equation}
 \rho_q(n_3)= \sum_{l_3}P_{l_3} (n_3)~ |n_3,l_3)(n_3,l_3|,~~~~\sum_{l_3}P_{l_3}=1,~~\forall ~n_3. \label{mixedst}
\end{equation}
Clearly $P_{l_3}$ are probabilities that are position-dependent. As mentioned in \cite{b1}, the distance formula (\ref{opdis}) will yield the true Connes' distance between the mixed states for which the probabilities $P_{l_3}$ are position independent.

However, instead of using the operator norm to calculate the infinitesimal distance, one can use trace norm which will give the closely related distance function 
\begin{equation}
 \tilde{d}(\rho_q(n_3+1, \rho_q(n_3))= \frac{\text{tr}_c(d\rho_q)^2}{\lVert [\mathcal{D},\pi(d\rho_q)]\rVert_{\text{tr}}}. \label{trdis}
\end{equation}
This distance given by (\ref{trdis}) will expectedly be different from the Connes infinitesimal distance given by (\ref{opdis}) by a numerical factor only so we can employ (\ref{trdis}) instead of (\ref{opdis}) for computational simplicity.

Now, introducing $d\rho_q(n_3+1,n_3)= \rho_q (n_3+1) -\rho_q(n_3)$, we can compute the closely related distance function between the mixed states on the subspace $\mathcal{H}_n$ of quantum Hilbert space using the formula (\ref{trdis}). After the straightforward computation, we get
\begin{equation}
 \text{tr}_q(d\rho_q(n_3+1,n_3))^2 = \sum_{l_3} [P^2_{l_3}(n_3+1) + P^2_{l_3}(n_3)],
\end{equation}
and 
\begin{eqnarray}
\nonumber \lVert [\mathcal{D},\pi(d\rho_q(n_3+1,n_3))]\rVert_{\text{tr}}=\frac{2}{\lambda r} \times~~~~~~~~~~~~~~~~~~~~~~~~~~~~~~~~~~~~~~~~~~~~~~~~~~~~~~~~~~~~~~~~~~~~~~~~~~~~~~~~\\
 \sqrt{\sum_{l_3}[P^2_{l_3}(n_3+1)\{n(n+1)-(n_3+1)^2\} + P^2_{l_3}(n_3)\{n(n+1)-n_3^2\}+
  P_{l_3}(n_3+1)P_{l_3}(n_3)\{n(n+1)-n_3(n_3+1)\}]},
\end{eqnarray}
so that we obtain the distance function as
\begin{eqnarray}
 \nonumber \tilde{d}(n_3+1,n_3) = \frac{\lambda r}{2}\times ~~~~~~~~~~~~~~~~~~~~~~~~~~~~~~~~~~~~~~~~~~~~~~~~~~~~~~~~~~~~~~~~~~~~~~~~~~~~~~~~~~\\
 \frac{\sum_{l_3} [P^2_{l_3}(n_3+1) + P^2_{l_3}(n_3)]}{\sqrt{\sum_{l_3}[P^2_{l_3}(n_3+1)\{n(n+1)-(n_3+1)^2\} + P^2_{l_3}(n_3)\{n(n+1)-n_3^2\}+
  P_{l_3}(n_3+1)P_{l_3}(n_3)\{n(n+1)-n_3(n_3+1)\}]}}. \label{dis.tr.}
\end{eqnarray}
Clearly, the distance depends upon the probabilities which shows the connection between geometry and statistics.
Proceeding in the same way as \cite{b1}, we can take two choices of probability distribution: one that minimize the distance between two generalized points and another that maximize the local entropy, while fixing the local average energy.

Let us consider the first choice. Since we have the infinitesimal distance between the mixed states, we can define the distance between two generalized points $n_i$ and $n_f$ on $\mathcal{H}_n$ as
\begin{equation}
 \tilde{d}(n_f,n_i)=\sum_{n_3=n_i}^{n_f-1}\tilde{d}(n_3+1,n_3).
\end{equation}
After long computation, we obtain that the probabilities that minimize the distance must satisfy
\begin{equation}
\Delta ~P_{l_3} = 2\alpha, ~~~~~~\forall~~l_3,  \label{mincond}
\end{equation}
where
\begin{equation}
\Delta=\begin{pmatrix}
          a(n_i) & b(n_i) & 0 & . & . \\
          b(n_i) & a(n_i+1) & b(n_i+1) & 0 & . \\
          0 & . & . & . & .  \\
          . & . & . & . & .  \\
          . & 0 & b(n_f-2) & a(n_f-1) & b(n_f-1)\\
          . & . & 0 & b(n_f-1) & a(n_f)
         \end{pmatrix};~~~ P_{l_3}=\begin{pmatrix}
          P_{l_3}(n_i)~~~~~~\\
          P_{l_3}(n_i+1)\\
          .\\
          .\\
          P_{l_3}(n_f-1)\\
          P_{l_3}(n_f)~~~~~~~
         \end{pmatrix};~~~\alpha=\begin{pmatrix}
          \alpha(n_i)~~~~~~\\
          \alpha(n_i+1)\\
          .\\
          .\\
          \alpha(n_f-1)\\
          \alpha(n_f)~~~~~~~
         \end{pmatrix}.
\end{equation}
Here, $\alpha(n_3)$ ($n_3$ taking value from $n_i$ to $n_f-1$)  are the Lagrange multipliers imposing the constraints that the probabilities sum to 1 and the matrix elements of $\Delta$ are given by
\begin{eqnarray}
\nonumber a(n_3)&=&\lambda\sqrt{n(n+1)}[g(n_3)+g(n_3-1) - \{n(n+1)-n_3^2\}\{f(n_3)+f(n_3-1)\}],\\
\nonumber b(n_3)&=&-\lambda\sqrt{n(n+1)}[\{n(n+1)-n_3(n_3+1)\}f(n_3)], 
\end{eqnarray}
with $f(n_3)$ and $g(n_3)$ given by
\begin{equation}
 f(n_3)= \frac{\frac{1}{2}\sum_{l_3} [P^2_{l_3}(n_3+1) + P^2_{l_3}(n_3)]}{\left[\sum_{l_3}[P^2_{l_3}(n_3+1)\{\big(\frac{r}{\lambda}\big)^2-(n_3+1)^2\} + P^2_{l_3}(n_3)(\big(\frac{r}{\lambda}\big)^2-n_3^2)+
  P_{l_3}(n_3+1)P_{l_3}(n_3)\{\big(\frac{r}{\lambda}\big)^2-n_3(n_3+1)\}]\right]^\frac{3}{2}},\label{f}
\end{equation}

\begin{equation}
 g(n_3)=\frac{1}{\sqrt{\sum_{l_3}[P^2_{l_3}(n_3+1)\{\big(\frac{r}{\lambda}\big)^2-(n_3+1)^2\} + P^2_{l_3}(n_3)(\big(\frac{r}{\lambda}\big)^2-n_3^2)+
  P_{l_3}(n_3+1)P_{l_3}(n_3)\{\big(\frac{r}{\lambda}\big)^2-n_3(n_3+1)\}]}}. \label{g}
\end{equation}
Here ~$\lambda^2n(n+1)=r^2$ is used in the equations (\ref{f}) and (\ref{g}) to shorten the expressions. 

From equation (\ref{mincond}), we see that $P_{l_3}$ is independent of $l_3$ since both $\Delta$ and $\alpha$ are independent of $l_3$ so that we get
\begin{equation}
 \sum_{l_3=-n}^n P_{l_3}(n_3)=1 \Rightarrow P_{l_3}(n_3)=\frac{1}{(2n+1)}.
\end{equation}
Substituting this in the equation (\ref{dis.tr.}), we get the distance function as
\begin{equation}
 \tilde{d}(n_3+1,n_3)=\frac{1}{\sqrt{(2n+1)}}\frac{\lambda\sqrt{n(n+1)}}{\sqrt{3\{n(n+1)-n_3(n_3+1)-\frac{1}{3}\}}}.
\end{equation}
This distance differs from the true Connes infinitesimal distance just by a numerical factor resulting from the use of the trace instead of operator norm.

Let us consider the second choice where we introduce a local entropy as
\begin{equation}
 S(n_3)=\sum_{l_3} P_{l_3}(n_3) ~ \log P_{l_3}(n_3),
\end{equation}
with the further condition that ~ $\sum_{l_3} P_{l_3}(n_3) E_{l_3}= E(n_3)$ ~ in addition to ~ $\sum_{l_3} P_{l_3}(n_3)=1$. After maximizing the local entropy, we get 
\begin{equation}
 P_{l_3}(n_3) = \frac{e^{-\beta(n_3) E_{l_3}}}{\sum_{l_3}~ e^{-\beta(n_3) E_{l_3}}} = \frac{e^{-\beta(n_3) E_{l_3}}}{Z(\beta(n_3))},\label{ent.pro.}
\end{equation}
where  $\beta(n_3)$  is the local inverse temperature introduced as a Lagrange multiplier imposing the local energy constraint and ~ $Z(\beta(n_3))=\sum_{l_3}~ e^{-\beta(n_3) E_{l_3}}$ ~ is the partition function.

If we take the local average energy and so the temperature to be independent of ~$n_3$,~ then putting (\ref{ent.pro.}) in (\ref{dis.tr.}), we get the distance function as
\begin{equation}
\tilde{d}(n_3+1,n_3) = \frac{\sqrt{Z(2\beta)}}{Z(\beta)}\frac{\lambda\sqrt{n(n+1)}}{\sqrt{3\{n(n+1)-n_3(n_3+1)-\frac{1}{3}\}}}. 
\end{equation}
This clear shows the connection between the distance and partition function describing the statistical properties of a system with quantum states given by (\ref{mixedst}) in thermal equilibrium. However, in this case the distance decreases as the temperature increases since the value of the factor ~$\frac{\sqrt{Z(2\beta)}}{Z(\beta)}$~ lies within ~$0$ to $1$.

\section{Conclusion}

The Hilbert-Schmidt operatorial formulation of the non-commutative quantum mechanics on the fuzzy space of the Lie algebra type of non-commutativity has been revisited and introducing the appropriate spectral triple on a particular fuzzy sphere indexed by $n$, the infinitesimal Connes distance has been calculated in both classical configuration space and quantum Hilbert space using the same prescription given in \cite{b1}. The connection between the geometry of the quantum Hilbert space and the statistical properties of the quantum system associated with the fuzzy sphere is shown by computing the infinitesimal distance function between the mixed states of the quantum Hilbert space.

\section*{Acknowledgements}
Shivraj Prajapat would like to thank the Department of Science and Technology (DST), Government of India, for providing financial support through INSPIRE fellowship and also the authorities of S N Bose Centre for their kind hospitality during the course of this work.

\section*{Appendices}
\subsection*{A1}
Under an infinitesimal $SU(2)$ transformation, the spinor  $\chi$ transform as 
\begin{eqnarray}
 \chi_\alpha\rightarrow\chi'_\alpha = \chi_\alpha+(i/2)\varepsilon^{i}\sigma^{i}_{\alpha\beta}\chi_{\beta}.
\end{eqnarray}
Taking each component $\Psi_{\alpha}$ of the section of the spinor bundle over $S^{3}$ to transform as a scalar,
\begin{eqnarray}
 \Psi_{\alpha}(\chi_{\alpha},\chi^{*}_{\alpha})\rightarrow \Psi'_{\alpha}(\chi'_{\alpha},\chi^{'*}_{\alpha}) = \Psi_{\alpha}(\chi_{\alpha},\chi^{*}_{\alpha}), \label{a1}
\end{eqnarray}
we get
\begin{eqnarray}
 \Psi'_{\alpha}(\chi_{\alpha},\chi^{*}_{\alpha}) +i\varepsilon^{i}\frac{1}{2}[\chi_{\beta}(\sigma^{i})_{\alpha\beta}\partial_{\alpha}-\chi^{*}_{\beta}(\sigma^{i})^{*}_{\alpha\beta}\partial^{*}_{\alpha}]\Psi_{\alpha} = \Psi_{\alpha}(\chi_{\alpha},\chi^{*}_{\alpha}),
\end{eqnarray}
so that 
\begin{eqnarray}
 \delta\Psi_{\alpha} = \Psi'_{\alpha}(\chi_{\alpha},\chi^{*}_{\alpha})-\Psi_{\alpha}(\chi_{\alpha},\chi^{*}_{\alpha}) = -i\varepsilon^{i}J_{i}\Psi_{\alpha},
\end{eqnarray}
where $J_{i}$ representing the ``orbital'' part of the rotation generator is given by the following differential operator
\begin{eqnarray}
J_{i} = \frac{1}{2}[\chi_{\beta}(\sigma^{i})_{\alpha\beta}\partial_{\alpha}-\chi^{*}_{\beta}(\sigma^{i})^{*}_{\alpha\beta}\partial^{*}_{\alpha}].
\end{eqnarray}
Of course, to obtain the complete rotational generator, we need to consider the spinorial transformation properties:
\begin{equation}
 \Psi_\alpha(\chi_\alpha,\chi^*_\alpha) \rightarrow \Psi'_\alpha(\chi'_\alpha,\chi'^*_\alpha) = U_{\alpha\beta}\Psi_\beta(\chi_\alpha,\chi^*_\alpha);~~~U\in SU(2),
\end{equation}
rather than (\ref{a1}). Again considering an infinitesimal transformation, one can gets an additional ``spin'' contribution. However, this is not required here.

\subsection*{A2}
 Since $S^{3}$ is the $U(1)$ bundle over $CP^{1}\sim S^{2}$,  there exist a natural $U(1)$ connection (gauge field) over $S^{2}$.  We would like to reproduce the computation of this connection term $A_{\mu}$ on the $S^{2}$ manifold using Atiyah's method \cite{b13}.  \\
 We shall first discuss the general construction and then the special case of $CP^{1}.$ Consider a vector bundle $E$ over the base space $M$ consists of a family of vector space $E_{y}$ parametrized by points $y \in M$. Also, let $E$ be a sub-bundle of the trivial bundle $M\times R^{N}$, such that $E_{y}$ can be embedded  as  a vector subspace in $R^{N}$ and any section $f(y)$ of $E$ taking its values in $E_{y}$ can be thought of as a function taking value in $R^{N}$. The partial derivative of $f$, may not take value in $E_{y}$ . The projection of the ordinary derivatives in to $E_{y}$ defines the covariant derivative on $E$, 
\begin{eqnarray}
 \nabla f = Pdf,~~~~~~~ P- \text{projection operator}.
\end{eqnarray}
If $E$ is the tangent bundle over $M$ and $P$ be the orthogonal projection, we get Levi-Civita connection of the Reimannian geometry. Choosing an orthogonal gauge/local frame for the bundle $E$ gives the linear maps $R^{n}\rightarrow E_{y}$ which are isomorphisms preserving orthogonality. Composing these isomorphisms with the continuous embedding of $E_{y}$ in $R^{N}$, we can write $U_{y}:R^{n}\rightarrow R^{N}$. The $U_{y}U^{\dagger}_{y} = P_{y}$ is then the projection operator and projects orthogonal elements in $R^{N}$ on to $E_{y}$. To calculate the covariant derivative $\nabla$ in the gauge '$U$' we put $f=Ug$, where $g$ is function on $M$ which takes values in $R^{n}.$
\begin{eqnarray}
 \nabla(Ug) = Pd(Ug) = UU^{\dagger}d(Ug) = U[dg+i(-iU^{\dagger}dU)g],
\end{eqnarray}
which shows that the gauge field is 
\begin{eqnarray}
 A= -iU^{\dagger}dU. \label{bh3}
\end{eqnarray}
To determine the $U(1)$ gauge field over $CP^{1}$, consider the tautological line bundle over $CP^{1}$. Now with  the gauge $z_{1} = \bar{z}_{1}$  the complex doublet $z =\left(
\begin{array}{c}
1/\sqrt{1+\bar{\rho}\rho}\\
\rho/\sqrt{1+\bar{\rho}\rho}\\
\end{array}
\right)$  represent a unit vector in $C^{2}$ and multiplication by the nonzero complex numbers $\lambda$ generates a whole comlex line through this point in $CP^{1}$ and defines section/gauge in this line bundle over $S^{2}$. The $U_{y}$ here is given by
\begin{eqnarray}
U_{y} = \left(
\begin{array}{c}
z_{1}\\
z_{2}\\
\end{array}
\right) = \left(
\begin{array}{c}
1/\sqrt{1+\bar{\rho}\rho}\\
\rho/\sqrt{1+\bar{\rho}\rho}\\
\end{array}
\right) 
\end{eqnarray}
and one gets, using (\ref{bh3}),
\begin{eqnarray}
 A = -iZ^{\dagger}dZ.   \label{io1}
\end{eqnarray}
For any other $U(1)$ bundle over $CP^1\sim S^2$ with the Chern class '$k$', one will get an additional factor of '$k$'.

\subsection*{A3}
 Using the $\chi$ parametrization  (\ref{f1},\ref{rr2}) one can obtained the metric over $S^{3}$, as induced from the flat metric on $C^{2}_{0}$, given as 
\begin{eqnarray}
 ds^{2} = d\chi^{\dagger}d\chi.
\end{eqnarray}
This is obtained as
\begin{eqnarray}
ds^{2} = d\chi^{\dagger}d\chi =\frac{1}{4} [d\theta^{2}+d\varphi^{2}+d\psi^{2}+2\cos\theta d\varphi d\psi].
\end{eqnarray}
Identifying this with the metric tensor through $ds^{2}= g_{ij}dx^{i}dx^{j}$  with  
 $(x^{1}=\theta, x^{2}=\varphi, x^{3} = \psi)$ one gets,
\begin{eqnarray}
 g_{ij} =  \left[ {\begin{array}{ccc}
 g(\partial_{\theta},\partial_{\theta}) & g(\partial_{\theta},\partial_{\varphi})  &  g(\partial_{\theta},\partial_{\psi})\\
 g(\partial_{\varphi},\partial_{\theta}) &  g(\partial_{\varphi},\partial_{\varphi}) &   g(\partial_{\varphi},\partial_{\psi})\\ 
 g(\partial_{\psi},\partial_{\theta}) &  g(\partial_{\psi},\partial_{\varphi}) &  g(\partial_{\psi},\partial_{\psi})
\end{array} } \right] = \frac{1}{4}\left[ {\begin{array}{ccc}
 1 & 0 & 0\\
0 & 1 & \cos\theta \\ 
0 & \cos\theta & 1
\end{array} } \right].
\end{eqnarray}
Using this the inner product between $J_{i}$  (\ref{c15}), (\ref{shiv})  over $S^{3}$ can be easily computed to get 
\begin{eqnarray}
 g(J_{i},J_{j})  = \frac{1}{4}\delta_{ij}.
\end{eqnarray}

\end{document}